\documentclass[fleqn,usenatbib]{mnras}

\usepackage{newtxtext,newtxmath}

\usepackage[T1]{fontenc}
\usepackage{ulem}
\usepackage{comment}

\DeclareRobustCommand{\VAN}[3]{#2}
\let\VANthebibliography\thebibliography
\def\thebibliography{\DeclareRobustCommand{\VAN}[3]{##3}\VANthebibliography}

\usepackage{graphicx}	
\usepackage{amsmath}	

\usepackage{amssymb}	

\newcommand{\beq}{\begin{equation}}
\newcommand{\eneq}{\end{equation}}
\newcommand{\becase}{\begin{cases}}
\newcommand{\encase}{\end{cases}}
\newcommand{\bet}{\begin{table}}
\newcommand{\ent}{\end{table}}

\title[N-body Moon Formation]{An N-body population synthesis framework for the formation of moons around Jupiter-like planets}

\author[Cilibrasi et al.]{
Cilibrasi, M.$^{1,2}$\thanks{E-mail: marco.cilibrasi@uzh.ch},
Szul{\'a}gyi, J.$^{2,1}$,
Grimm, S. L.$^{3}$,
and Mayer, L.$^{1}$
\\
$^{1}$Institute for Computational Science, University of Z\"urich, Winterthurerstrasse 190, CH-8057 Z\"urich, Switzerland\\
$^{2}$ETH Z\"urich, Department of Physics, Wolfgang-Pauli-Strasse 27, CH-8093, Z\"urich, Switzerland\\
$^{3}$Center for Space and Habitability, Universität Bern, Gesellschaftsstrasse 6, CH-3012 Bern, Switzerland
}

\date{Accepted XXX. Received YYY; in original form ZZZ}

\pubyear{2020}

\begin{document}
\label{firstpage}
\pagerange{\pageref{firstpage}--\pageref{lastpage}}
\maketitle

\begin{abstract}
The moons of giant planets are believed to form in situ in Circumplanetary Discs (CPDs). Here we present an N-body population synthesis framework for satellite formation around a Jupiter-like planet, in which the dust-to-gas ratio, the accretion rate of solids from the Protoplanetary Disc, the number, and the initial positions of protosatellites were randomly chosen from realistic distributions. The disc properties were from 3D radiative simulations sampled in 1D and 2D grids and evolved semi-analytically with time. The N-body satellitesimals accreted mass from the solid component of the disc, interacted gravitationally with each other, experienced close-encounters, both scattering and colliding.
With this improved modeling, we found that only about $15\%$ of the resulting population is more massive than the Galilean one, causing migration rates to be low and resonant captures to be uncommon. In 10\% of the cases, moons are engulfed by the planet, and 1\% of the satellite-systems lose at least 1 Earth-mass into the planet, contributing only in a minor part to the giant planet's envelope's heavy element content. We examined the differences in outcome between the 1D and 2D disc models and used machine learning techniques (Randomized Dependence Coefficient together with t-SNE) to compare our population with the Galilean system. Detecting our population around known transiting Jupiter-like planets via transits and TTVs would be challenging, but $14\%$ of the moons could be spotted with an instrumental transit sensitivity of $10^{-5}$.
\end{abstract}

\begin{keywords}
planets and satellites, formation - planets and satellites, gaseous planets - planets and satellites, general
\end{keywords}



\section{Introduction}

Two main theories of giant planet formation in Protoplanetary Discs (PPDs) have been developed in the last few decades, the Core Accretion Model (CA; \citealt{Pollack96}) and the Gravitational Instability Model (GI; \citealt{Boss97,Durisen07}). 
In both these scenarios, in the final phase of the formation process, a  Circumplanetary Disc (CPD) made of gas and dust develops around the nascent planet \citep[e.g.][]{Kley99,Lubow99,Canup02,Ayliffe09,Shabram13,Szulagyi14,Szulagyi17gap}.
CPD structure and evolution have been investigated mainly thanks to semi-analytical models \citep{Alibert05, Ward10} and hydrodynamical numerical simulations \citep[e.g.][]{Ayliffe09, Shabram13, Szulagyi14, Szulagyi17gap}. Even though CPDs can be seen as small Protoplanetary discs around a planet, the most important difference is that PPDs are independent structures which, at least in their late stages, can be assumed to be isolated from their parent star forming clouds, while CPDs are continuously fed by a flow of material (gas and possibly dust) from the PPDs, coming from high latitudes and accreting vertically \citep{Tanigawa12, Szulagyi14}.

Circumplanetary Discs are considered the birth places for satellite systems around giant planets \citep{Canup02}. As we see in our Solar System, giant planets often have systems of regular satellites, massive enough to be spherical, with low eccentricities and inclinations, similar to a scaled-down version of a planetary system. After the first indirect evidence of V1400 Centauri \citep{Kenworthy15}, where gaps in a detected ring system were supposed to be caused by exomoons, the first exomoon candidate has been spotted also in an exoplanetary system as a Neptune-size moon orbiting the 10 Jupiter masses planet Kepler-1625b \citep{Teachey18}. Due to the so-far gas only hydrodynamic simulations of the circumplanetary discs, we do not know yet what solids content they might have, or what grain-size distribution one can expect. It is known that the ratio between the total mass of moons around Jupiter, Saturn and Uranus and their planetary mass is about $2\times10^{-4}$, implying a minimum total gas mass of the CPD of $2\times10^{-2}$ planetary masses (computed with the interstellar medium value of dust-to-gas ratio: 1\%). This is called Minimum Mass Sub-nebula Model, or MMSM \citep{Mosqueira03}. Such a massive disc would imply some difficulties for satellite formation, such as too fast migration rates (i.e. all the moons can be lost by migrating into the planet) or too hot temperatures for ice to condense, but a continuous feeding from the PPD through the meridional circulation \citep{Szulagyi14} would allow a much lighter CPD to exist over time, while still able to form enough satellites, in what is usually called \textit{gas-starved} scenario \citep{Canup06}. The persistent flow of solids from the PPD could possibly allow the formation of more than one generation of satellites, migrating one after the other into the central planet, in the so-called \textit{sequential formation} scenario \citep{Canup02}.

There have been several studies of satellite formation in Circumplanetary Discs so far.  \cite{Fujii17} used a 1D analytical model for moon formation and migration in a CPD, and found that moons are often locked in resonant chains, in agreement with the case of the Galilean satellites, while \cite{Fujii20} also implemented an N-body integrator and managed to produce single-moon systems as Titan around Saturn. \cite{Heller15} studied the evolution of a CPD also taking into account different heating processes, focusing on the position of the ice-line and concluding that satellites should form in the very last phase of a giant planet formation, when the environment is sufficiently cold for icy satellite formation. 
Other works have been focusing on the accretion processes of satellites. Other than the satellitesimal accretion and dust accretion, the pebble accretion scenario has been investigated in the last few years. \cite{Shibaike19} showed that slow pebble accretion of planetesimals captured in CPDs would be able to prevent a too fast migration process and, given some fine-tuned halt conditions and physical constraints, to build Galilean-like systems in masses, positions (and resonances), and compositions. \cite{Ronnet20} showed that pebbles would not be able to flow from the PPD to the CPD because of the pressure gradient at the gap edge, while the smaller grains are captured by gas drag. This mechanism would ablate pebbles, but providing the circumplanetary disc with small dust grains, which then coagulate into pebbles inside the CPD. This pebble accretion scenario furthermore represent rapidly migrating satellites, which are piling up at the inner boundary of the CPD, creating resonant chains.

A common approach in studying satellite formation is the so-called \textit{population synthesis} method, that have been already widely applied to planet formation \citep{Ida04, Benz14, Mordasini18}. Using such an approach means running many thousands of individual simulations, varying initial conditions, for example the moonlets' initial positions, the disc dust-to-gas ratio, its viscosity, or lifetime. An early example of an application of this method to moon-formation comes from \cite{Sasaki10}, in which the authors numerically computed the viscous evolution of a CPD, following the prescriptions of \cite{Pringle81}, including a cavity between the disc and the planet, solid accretion onto the satellite-seeds and type I migration as in \cite{Tanaka02}. Their simulations resulted in systems of four-five satellites, often locked in resonances. A similar approach has been followed by \cite{Miguel16}, in which the CPD was modeled as a MMSM, with randomly choosing the initial positions of 20 seeds, their initial masses, and the dust-to-gas ratio of the disc. Their results show that the high density gas causes very rapid migration of satellitesimals. Nevertheless, the conditions in the disc also favor the formation and the survival of large satellites, but massive and far-away bodies are difficult to reproduce.
Another population synthesis approach was shown in \cite{Cilibrasi18}, where the CPD was an azimuthally and vertically integrated version of a 3D radiative hydrodynamics simulations of Jupiter's CPD \citep{Szulagyi17gap}. Up to 20'000 simulations were run with different dust-to-gas ratios, disc life-times and timescales of dust refilling from the PPD, showing a great occurrence of sequential formation and massive satellites in this 1D limit. Similar models have been also implemented to study satellite formation around ice giants in the Solar System \citep{Szulagyi18b} and around Jupiter-like exoplanets that are forming with Gravitational Instability scenario resulting in significantly different CPD characteristics \citep{Inderbitzi20}.

Another approach to study satellite formation is using N-body simulations. Instead of solving the motion of satellites and their migration in a 1D scenario as all previous examples, a proper N-body integrator can reach a much better accuracy in the interactions between satellites (including resonances), as in \cite{Moraes18}. Those authors simulated a MMSM disc, in which embryos could move in a 3D space, following gravity interactions, together with smaller satellitesimals, that were the main source of accretion. They run a few different systems with the MERCURY N-body integrator \citep{Chambers99}, showing the influence of the number of initial seeds, the temperature of the disc and its density on the final architecture of the systems.

In this paper, we present an N-body population synthesis framework for satellite formation around Jupiter-like planets, combining the advantages of the last two described approaches. In our model, satellites were treated as N-body particles in an integrator including close encounters \citep{Chambers99, Grimm14} and individual time-steps \citep{Saha94}, while more than two thousand simulations were run with different initial conditions, i.e. with randomly choosing the dust-to-gas ratio, the solid accretion rate from the PPD and the initial positions and number of the seeds. We also included the accretion onto the moonlets from a solids disc, and the interaction between the seeds and the disc itself, resulting in migration, eccentricity and inclination damping \citep{Cresswell08}. Furthermore, following the approach of \cite{Cilibrasi18}, our model does not replicate a MMSM or gas-starved disc, but uses 3D thermo-hydrodynamical simulations from \cite{Szulagyi17gap} to build a 1D disc model, in which the N-body framework is embedded along with semi-analytical recipes to describe processes involved in the interaction between the satellitesimals and the surrounding CPD. We also tested the evolution of satellites, particularly their accretion, when the integrator is embedded in a 2D disc model, from \citet{Szulagyi17gap}. We compared the results of the 1D disc + 3D N-body approach, and the 2D disc + 3D N-body case. 


\section{Methods}\label{Methods}

\begin{figure}
\includegraphics[width=\columnwidth]{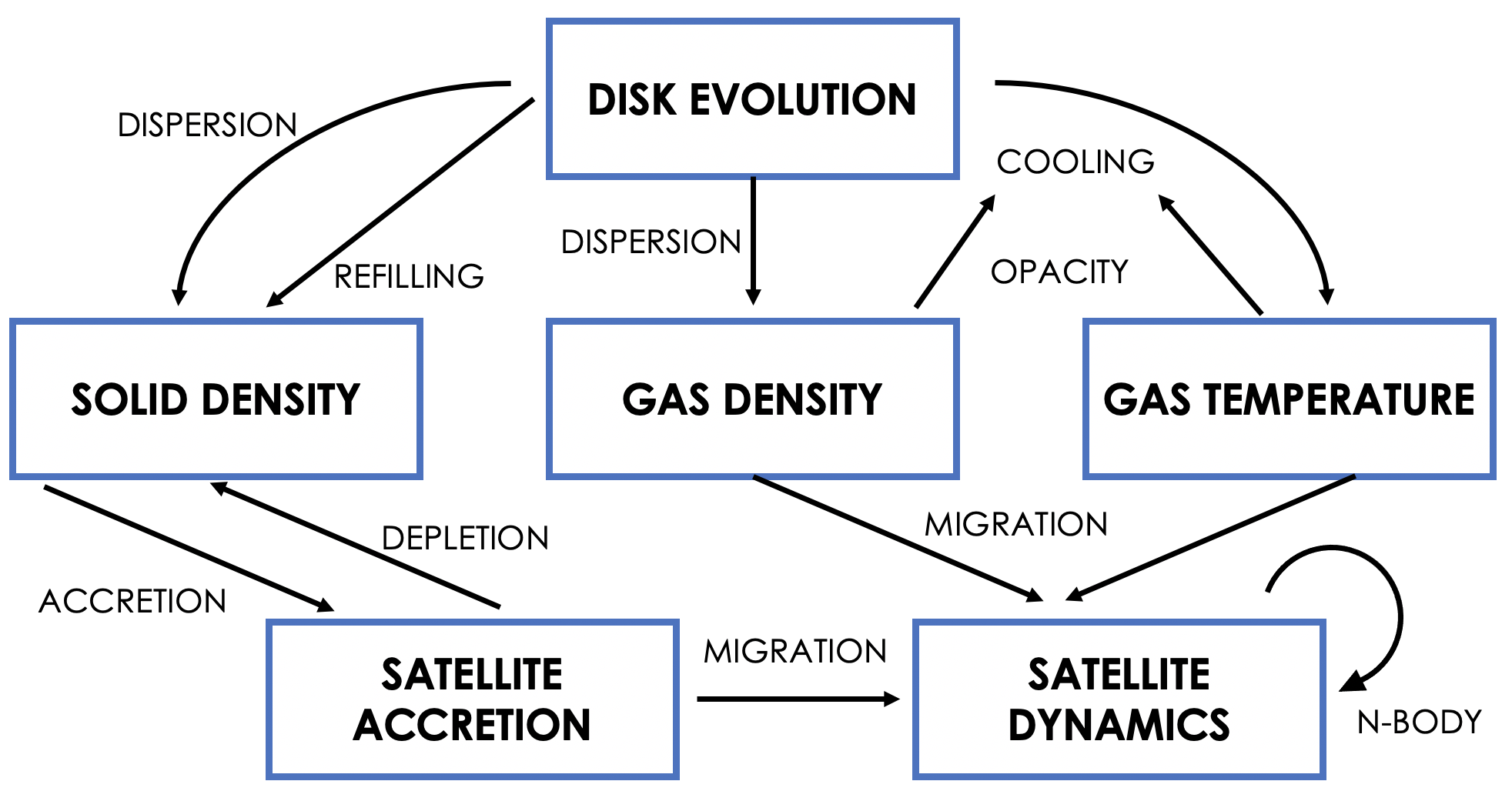}
\caption{The picture shows how the processes that are described in Section \ref{Methods} depend on each other. For example, the gas density, temperature and mass (here implicitly considered within the satellite accretion processes) have an effect on satellite migration, or also the solid density determines the satellite accretion rate, that in turn decreases the solid density.}\label{ModelScheme}
\end{figure}

The model presented in this paper is a semi-analytical model, consisting of a CPD in which satellites are treated as growing embryos that migrate through the disc and interact with each other. Since migration and accretion of protosatellites depend on the disc properties, such as gas temperature and density, and solids density, the disc is also evolving, meaning that it is dissipating and cooling, while the solid profile is evolving accordingly to satellite accretion. A scheme of how these processes and properties depend on each other is provided in Figure \ref{ModelScheme}.
In the model, the units are always $R_J$ for distances, $M_J$ for masses, $K$ for temperatures and $yr$ for times.

\subsection{Disc Model}\label{Disc Model}

\subsubsection{Disc Structure}\label{Disc Structure}

The circumplanetary disc has been modelled using the results of the simulations described by \cite{Szulagyi17gap} for Jupiter's circumplanetary disc. These simulations were the base also for our previous work \citep{Cilibrasi18}, but this time using the CPD model with the coldest planetary temperature (1000 K instead of 2000 K as in \citealt{Cilibrasi18}, in order to have a CPD that is forming later in the planet formation process and that could favour the formation of icy satellites) among the ones presented in the \cite{Szulagyi17gap}.
As in \cite{Cilibrasi18}, the CPD has gas and solid density profiles, a temperature profile and the same parameters as in \cite{Szulagyi17gap}: $\alpha = 0.004$ for viscosity \citep{Shakura73}, the heat capacity ratio $\gamma = 1.4$ (molecular hydrogen), and the mean molecular weight $\mu = 2.37$. The parameter $\alpha$ has a role only in type II migration, as explained in Section \ref{Migration and Damping}. The planet is similarly assumed to be a Jupiter-equivalent at 5.2 AU, with $M=1M_J$ and $R=1R_J$.

In this model, the radial profiles of the CPD are built on a 1D radial grid, ranging logarithmically from $1 R_J$ to $500 R_J$ (about $2/3$ of the Hill radius) with 1000 cells in total. A cavity between the planet’s surface and the disc is not considered in the model, similarly to the original hydrodynamical simulations. This is because the magnetic field of the young Jupiter and the ionization of the CPD might have not been strong enough to produce such a cavity at the time of formation \citep{Owen16}.
The initial profiles of the disc were taken as interpolations of the values coming from the hydrodynamic simulations, averaged both azimuthally and vertically in the case of the 1D density profile, or taken in the mid-plane as in the case of the temperature. Averaging azimuthally greatly decreases the computational expense and introduces a negligible approximation.

In fact, as explained in Section \ref{Satellite-disc interactions}, the disc structure interacts with the forming satellites via two processes: the gas component via migration and the solids via accretion.
The key parameters on which migration depends are the gas surface density $\Sigma$ and the local aspect ratio of the disc $H = h/r$. These two quantities are defined as follows
\begin{equation}
    \begin{split}
    \Sigma(r) = \int_{-\infty}^{+\infty}\rho(r, z) dz\\
    H(r) = \frac{h(r)}{r} = \frac{c_s(r)}{\Omega_K(r)r}
    \end{split}
\end{equation}
where $\rho$ is the gas volume density, $c_s = \sqrt{\gamma\frac{P}{\rho}} = \sqrt{\gamma\frac{k_B T}{\mu}}$ is the sound speed in the mid-plane (with $k_B$ being the Boltzmann constant) and $\Omega_K$ is the Keplerian angular velocity. 
Furthermore, migration is always much slower than orbital motion (depending also on the mass of the satellite, the migration timescale is from $10^8$ to $10^5$ times slower than the orbital period, as shown in Section \ref{Migration formulas}). As a result, since the orbital velocity of gas is sub-Keplerian \citep{Pringle81}, then any effect of azimuthal variations of surface density and aspect ratio on migration are averaged in time and negligible in the model. Unfortunately, this does not apply to the solids, as its orbital velocity is considered to be Keplerian, but in Section \ref{1D vs 2D Accretion} we show that the accretion process is also azimuthally averaged, this time by the differential velocity of the solids within the feeding zone of a satellite.

The dust-to-gas ratio ($Z$) of CPDs is still unknown. Therefore, we choose the dust-to-gas ratio of the disc as one of the initial parameters that are varied between the different runs of population synthesis (see in Section \ref{Results}).

\begin{figure}
\includegraphics[width=\columnwidth]{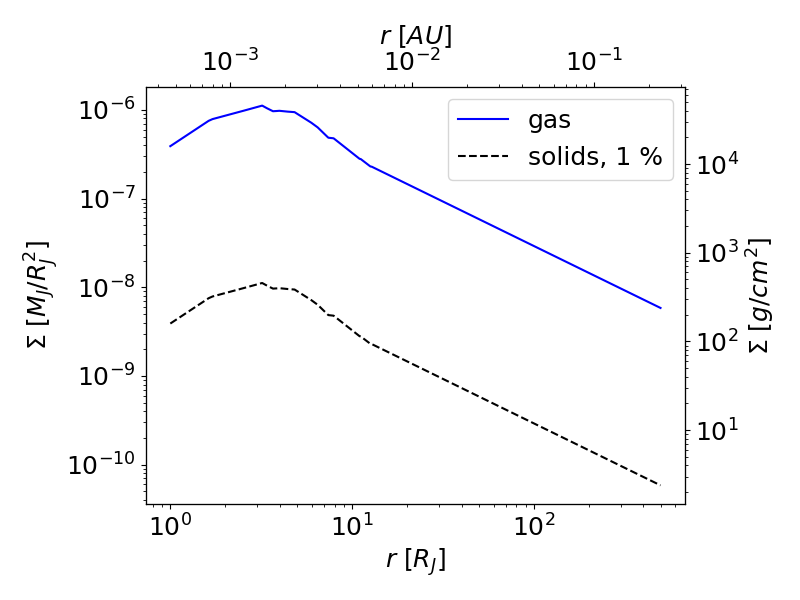}
\caption{Gas (blue) and solid (black) density profiles of the circumplanetary disc at the beginning of the simulations. The dust-to-gas ratio here is chosen to be 1\% as an example, but varied between the different runs.}\label{density_profile}
\end{figure}

\begin{figure}
\includegraphics[width=\columnwidth]{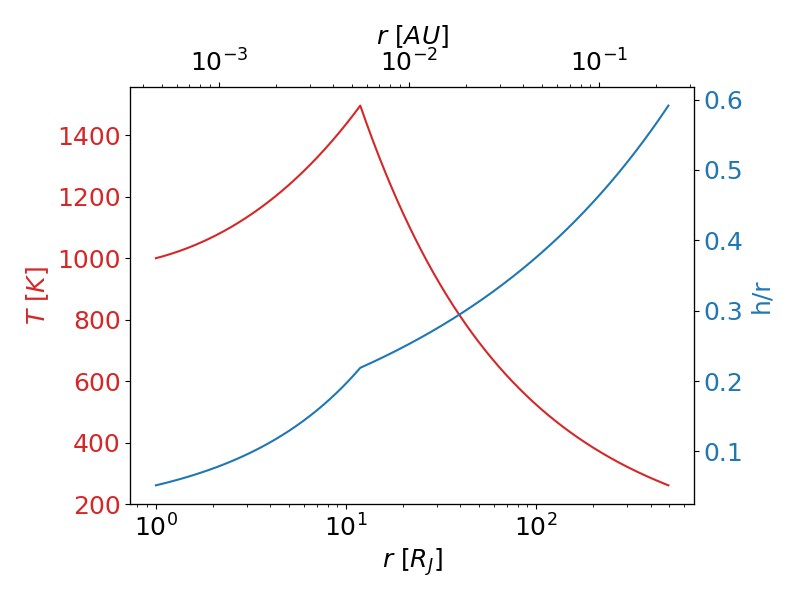}
\caption{Temperature (red) and aspect ratio (blue) initial profiles. The plot clearly shows the thickness of the CPD.}\label{temperature_profile}
\end{figure}

The initial disc density and temperature profiles are shown in Figure \ref{density_profile} and \ref{temperature_profile}, respectively. With these profiles, the total mass of the CPD is found to be $\simeq 9\times 10^{-3} M_J$, and, in the outer part, the thickness of the disc is about one order of magnitude bigger than the one usually assumed for protoplanetary discs ($\sim0.5$ versus $0.05$).

\subsubsection{Disc Evolution}\label{Disc Evolution}

The initial density and temperature profiles of the CPD are evolving in time in our semi-analytical runs, because the disc is dissipating and cooling. As in \cite{Cilibrasi18}, the gas density and temperature evolution is treated as an exponential law
\begin{equation}
\begin{split}
    &\Sigma(r, t) = \Sigma_0(r) e^{-t / t_{\text{disp}}}\\
    &T(r,t) = \left[T_0(r) - T_{\text{min}}\right]e^{-t / t_{\text{cool}}} + T_{\text{min}}
\end{split}
\end{equation}
where the dispersion time is taken to be $t_{\text{disp}} = 10^5$ years \citep{Fedele10}. In \cite{Cilibrasi18}, a distribution of possible dispersion timescales have been considered, but the effect of different values for this parameters has been shown to be negligible, especially regarding the architecture of the final satellite systems (mass and semi-major axis). For this reason, the smallest possible value has been chosen. That means that the disc is completely dissipated by the time the simulations end, i.e. after $10 t_{\text{disp}} = 10^6$ years. Consequently, the model is able to capture the system evolution even after the gas is completely dissipated (i.e. in the debris disc phase). The minimum background temperature at Jupiter location is assumed to be $T_{\text{min}} = 130 K$ (as in \citealt{Miguel16}) and the cooling time is calculated via radiative cooling (see \citealt{Wilkins12, Cilibrasi18}) with a timescale of $t_{\text{cool}}\simeq1.6\times10^5$ years.

The solid profile evolution is more complex. Overall, the solids are dissipating with the same time evolution as the gas component, but the interaction with satellites causes some other effects that are taken into account. First of all, as described in Section \ref{Accretion}, the satellites accrete mass from the solid component, with a certain accretion rate $\dot{M}$ within their feeding zone. Those cells that are part of the feeding zone will experience a density decrease that we call $\dot{\Sigma}_{\text{acc}}$. Furthermore, the same refilling mechanism used in \cite{Cilibrasi18} to model the solid accretion rate from the PPD to the CPD has been implemented, i.e. the solids coming from the PPD will replenish the gaps in the CPD created by solid accretion of the seeds in a timescale ($t_{\text{ref}}$). This effect tends to bring the density of solids to an equilibrium solution, that is decreasing exponentially. This timescale, as the dust-to-gas ratio of the disc, is an unknown parameter and will be treated as a free parameter in this model.

Mathematically, the equation for the evolution of solids that is solved in the model is the following:
\begin{equation}
    \frac{d\Sigma(r,t)}{dt} = - \frac{\Sigma(r,t)}{t_{\text{disp}}} - \dot{\Sigma}_{\text{acc}}(r,t) + \frac{\Sigma_0(r) e^{-t / t_{\text{disp}}} - \Sigma(r,t)}{t_{\text{ref}}}
\end{equation}
where the first term represents the exponential evolution, the second is the source term due to accretion, and the third term represents the refilling mechanism. With no accretion, the solution reduces to $\Sigma(r,t)=\Sigma_0(r) e^{-t / t_{\text{disp}}}$.

\subsection{N-body interactions}\label{N-body interactions}

In our model, protosatellites are treated as massive particles orbiting around the central planet, interacting with the CPD and with each other in 3D, using a 3D N-body integrator that we developed following the methods of GENGA \citep{Grimm14}. Here we give a brief description of the algorithm, all mathematical details are described in Appendix \ref{N-body integrator}, while a validation of the code, and its comparison with GENGA, are provided in Appendix \ref{N-body code validation}.

To begin with, each simulation starts with 10 to 20 embryos randomly distributed in the disc between 30 and 200 $R_J$, and the initial semi-major axis distribution is such that $\log_{10} a$ is uniform. These values have been chosen so that satellites are forming well within the Hill radius of Jupiter, but not too close to the planet, as it has been shown that the best satellite-forming location in CPDs should be at around $85 R_J$ \citep{Drazkowska18}. All embryos initially have 0 eccentricity and a random small inclination, distributed linearly between $-0.01$ and $0.01$, i.e. the typical values for the current Galilean satellites. Any lost satellite (because of collisions or ejections, as explained further in this section) is replaced with a new embryo in another random position in the disc, making the initial number of embryos also the maximum number of particles in the disc at a given moment. The initial mass of satellites is chosen to be $10^{-7}M_J$ and a new embryo is created only if the total mass of the solids present in the disc in that moment is at least 10 times the mass of the embryo itself. This is a way to mimic the continuous formation of seeds in a sequential formation scenario \citep{Ida10,Miguel16}.

Within the N-body integrator, satellites are considered as particles orbiting a central fixed particle with $M = 1M_J$. As the central particle is fixed, we consider it as an external potential, and all the sums in this Section will refer to the N satellites only. This simplifies the algorithm, since we do not need to take into account the effect of the planet movement in the disc, and it is reasonable, since the mass of the satellites is always significantly lower than the planetary mass (see in Section \ref{Retained and lost mass} for the total mass distribution).

In order to speed the code up, without losing accuracy, we implemented the individual time-step scheme described by \cite{Saha94}. In fact, in planet and satellite systems, the particles of an integrator just orbit around the central mass most of the time at Keplerian speed. Assigning longer time-steps to the satellites in the outer part of the system, we managed to minimize the running time, without losing any interaction.

This algorithm is built by splitting the Hamiltonian in its Keplerian and Interaction part (Appendix \ref{N-body integrator}). This splitting does not make sense any more when two satellite-particles get close to each other. In order to deal with it we also implemented the same close-encounter treatment used in GENGA \citep{Chambers99,Grimm14}. This consists in shifting the interaction between those particles to the Keplerian part. In order to be implemented, this method needs to redefine how the system's Hamiltonian is solved, but an appropriate synchronization method allows both individual time-steps and close-encounter treatment to co-exist.

\subsection{Collisions}\label{Collisions}

During the N-body integration, collisions are also possible. Assuming a density around the mean density of Galilean satellites ($\rho\simeq2.5g/cm^3$), a radius is assigned to each satellite. When the distance between two satellites is less than the sum of their radii, then they are considered colliding. Then, the two satellites are replaced by one, having the sum of the individual colliding moons' masses, situated in their common centre of mass and proceeding with the sum of the two momenta. In other words, the collisions are perfectly inelastic, meaning that the momentum is conserved, but the energy is not. In this model, we did not consider the possibility that satellites could also experience disruptive collisions, fragmentation or only partial merging. That would require a significantly more detailed modelling resolving their radii, which is beyond the scope of this paper. 

Satellites can also collide with the central planet (i.e. get engulfed by it). That happens when the radial distance of a satellite from the centre is less than $1R_J$. In this case the satellite is deleted from the simulation and its mass is considered as lost into the planet. The third possibility is that a satellite assumes an eccentricity that is $\ge1$. In this case the orbit is parabolic or hyperbolic, hence ejected from the system.

In the above three scenarios, when one satellite is lost, it is replaced by another embryo, randomly located in the disc following the same initial prescriptions regarding the initial position and mass, as described before.

\subsection{Satellite-disc interactions}\label{Satellite-disc interactions}

\subsubsection{Migration and Damping}\label{Migration and Damping}

While interacting with each other within the N-body integrator, satellites also interact with the disc itself. This interaction causes migration, i.e. a change in the semi-major axis of the orbit (usually inward), and eccentricity and inclination damping. In this model, these effects are implemented following the prescriptions by \cite{Cresswell08}.
These additional accelerations are included in the Interaction (Kick) part of the Hamiltonian as
\begin{equation}
    \begin{split}
        &\vec{a}_{\text{mig}} = -\frac{\vec{v}}{t_{\text{mig}}}\\
        &\vec{a}_{\text{ecc}} = -2\frac{v_r}{t_{\text{ecc}}}\hat{r}\\
        &\vec{a}_{\text{inc}} = -\frac{v_z}{t_{\text{inc}}}\hat{z}
    \end{split}
\end{equation}
and 

\begin{equation}
    \begin{split}
        &t_{mig} = \frac{2t_{\text{wave}}}{b_{\text{mig}}}\left(\frac{h}{r}\right)^2F_{\text{mig}}(e,i)\\
        &t_{\text{ecc}} = \frac{t_{\text{wave}}}{0.78}F_{\text{ecc}}(e,i)\\
        &t_{\text{inc}} = \frac{t_{\text{wave}}}{0.544}F_{\text{inc}}(e,i)
    \end{split}
\end{equation}
where $F_{\text{mig}}$, $F_{\text{ecc}}$, and $F_{\text{inc}}$ are functions of eccentricity and inclination defined in \cite{Cresswell08}, and $t_{\text{wave}} = \frac{M_J}{m}\frac{M_J}{\Sigma_{\text{gas}}a^2}\left(\frac{h}{r}\right)^4\Omega_K^{-1}$.
The parameter $b_{\text{mig}}$ for Type I migration can be found in several papers in literature, where it is calculated in different conditions \citep{Tanaka02, Dangelo10, Paardekooper10, Paardekooper11}. In this model, the calculations by \cite{Jimenez17} have been implemented, where the formula was derived from 3D isothermal and adiabatic simulations. It can be applied to low- and intermediate-mass planets in optically thick protoplanetary discs, or to protosatellites embedded in CPDs.
$b_{\text{mig}}$ is the final torque $\tau$ they derived divided by $\tau_0 = \Sigma_{\text{gas}}\Omega_K^2a^4\left(\frac{m}{M_J}\right)^2\left(\frac{h}{r}\right)^{-2}$.

As shown in \cite{Cilibrasi18}, Type II migration, when planets/moons open gaps in their gaseous discs, is not likely to happen with satellites forming within circumplanetary discs. For this reason, the effect of the choice for the parameter $\alpha$ is negligible in the model, as this parameter enters in the calculations only in this context. Nevertheless, we implemented Type II migration too in our model, using the gap-opening condition from \cite{Crida07} and the same scheme as \cite{Cilibrasi18}.

\subsubsection{Accretion}\label{Accretion}

The satellites accrete solid materials from the CPD, acquiring mass this way too. In particular, satellites that are moving across the disc will always accrete the solids from within their feeding zone. In this model, we assumed the solid component of the disc to be made of satellitesimals. Following \cite{Greenberg91}, the planetesimal solid accretion rate, applied to satellites, is:
\begin{equation}
\dot{M}_s=2\left(\frac{R}{a} \right)^{1/2} \Sigma_{\text{solid}} a^2 \left( \frac{M_s}{M_J}\right)^{1/2} \Omega_K
\end{equation}
and the feeding zone extends up to $2.3R_{\text{Hill}}$. 
When a satellite accretes, the cells within its feeding zone are going to have lower density of solids as a result. The same way, accretion happens as long as the cells within the feeding zone have enough mass to be accreted. The cells are then replenished by the refilling mechanism explained in Section \ref{Disc Evolution}.

\section{Results}\label{Results}

The satellite formation model described in Section \ref{Methods} was run following a population synthesis approach \citep{Ida04, Benz14, Mordasini18}. Three parameters were varied randomly with different values for each simulation, producing different initial conditions and therefore, different outcome. The three parameters were the dust-to gas ratio, the refilling timescale $t_{\text{ref}}$ and the number of initial satellite seeds.
The idea of populations synthesis is to statistically analyse the outcome of random initial parameters on the resulting population.

\begin{figure}
\includegraphics[width=\columnwidth]{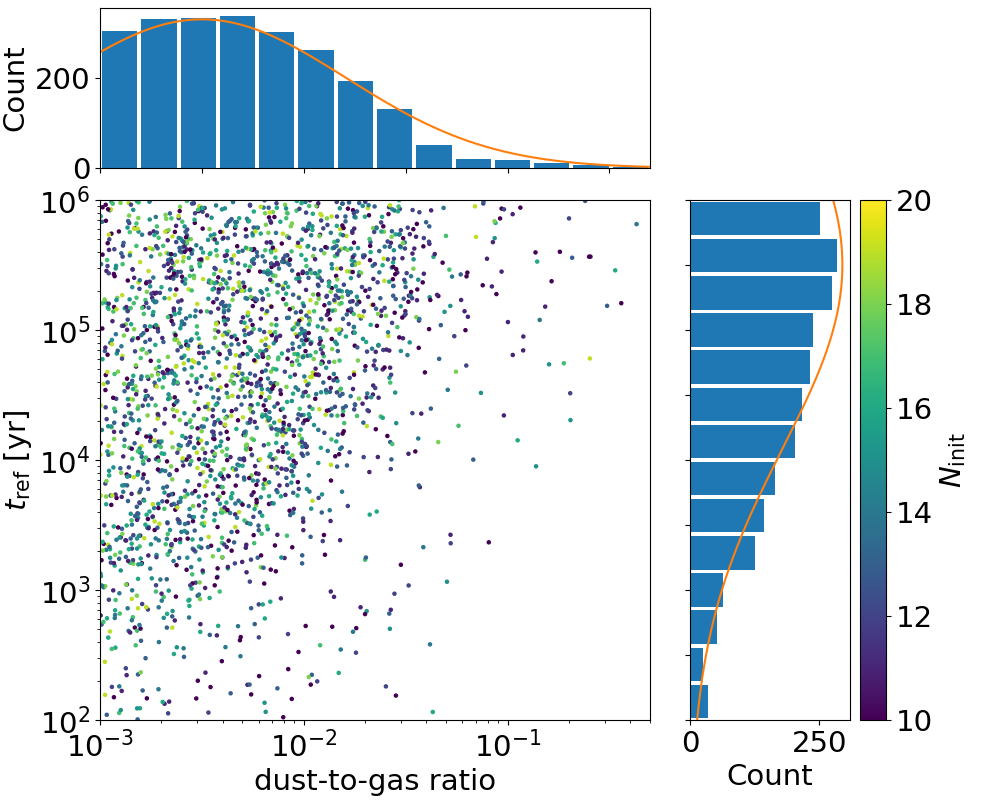}
\caption{Distribution of dust-to-gas ratios and refilling timescales from the protoplanetary disc in the population synthesis. The histograms represent the dust-to-gas ratio and refilling timescale distributions, while colors represent the initial number of seeds.}\label{parameter_distribution}
\end{figure}

In this case the ranges for the parameters have been chosen to be
\begin{equation}
    \begin{cases}
        \text{dust-to-gas ratio}\, (Z)\;\;\;\;\;&0.001-0.5\\
        t_{\text{ref}}\;\;\;\;\;&10^2-10^6\;\text{years}\\
        N_{\text{init}}\;\;\;\;\;&10-20
    \end{cases}
\end{equation}
The distribution shapes for these parameters were chosen in different ways. The initial number of seeds has been chosen uniformly in the range between 10 and 20. The other two parameters were distributed finding a compromise between the same logarithmic distributions used in \cite{Cilibrasi18} and the computational time. In fact, systems with a higher dust-to-gas ratio and a faster refilling, have more massive satellites and, consequently, the close-encounter integrator is much slower. 
The resulting distributions are shown in Figure \ref{parameter_distribution}. A total number of 2309 systems were simulated, enough to reach  convergence on the results. We tested higher simulation numbers too, but the shape and characteristic parameters (mean, variance) of the distributions did not change anymore above 2000 runs.


Once all the simulations have been completed, the idea of a population synthesis is to look at the statistics of the results, gaining a general understanding of how satellite systems around Jupiter-like planets might look like. On the other hand, since this analysis depends on the initial parameter distribution, another approach is looking at the dependence of the results on these parameters, allowing pattern identification and a deeper understanding of the significance of the results.

\subsection{Retained and lost mass}\label{Retained and lost mass}

\begin{figure}
\includegraphics[width=\columnwidth]{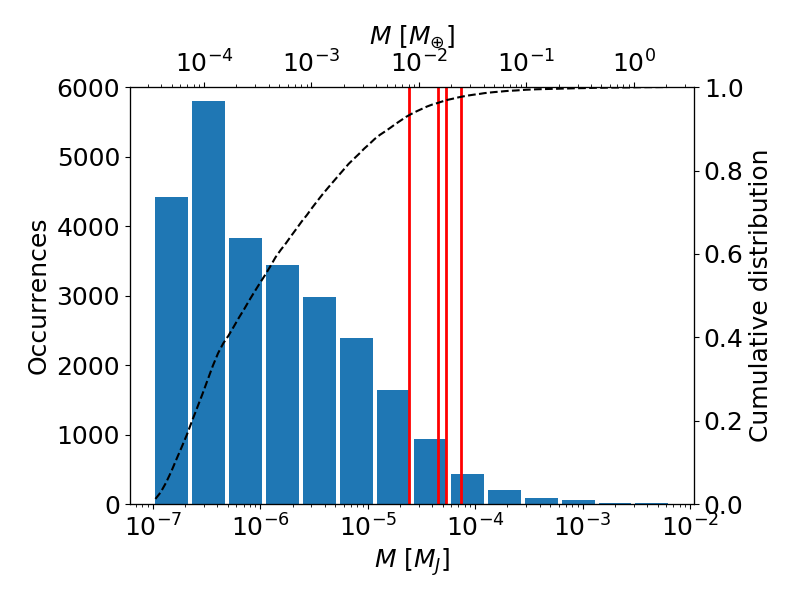}
\includegraphics[width=\columnwidth]{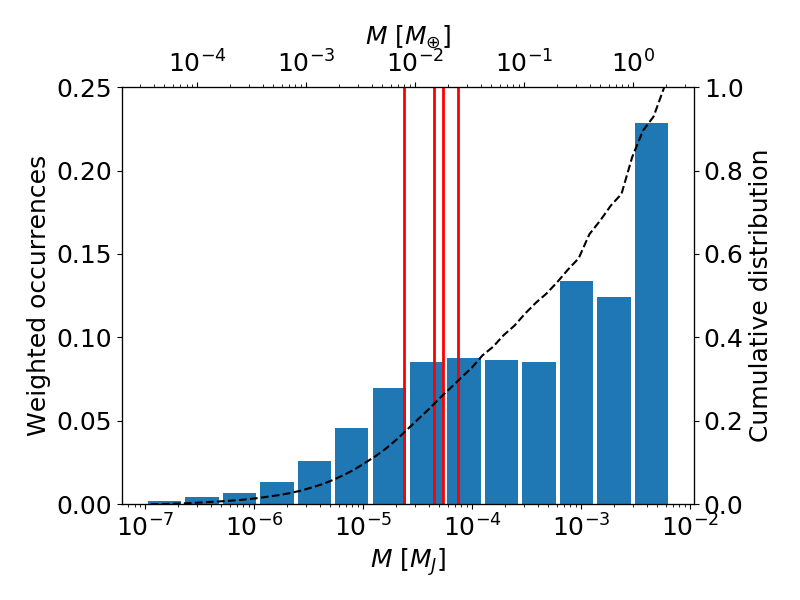}
\caption{The two figures show the distribution of the mass of survived satellites compared to the Galilean ones (red lines). In the second case, the distribution has been weighted by the mass of satellites itself, so that the sum of the values of all the bins equals 1. This highlights that massive satellites are lower in number, but more important in terms of mass budget. In both cases the dashed lines represent the cumulative distribution of the masses.}\label{single_mass_distribution}
\end{figure}

\begin{figure}
\includegraphics[width=\columnwidth]{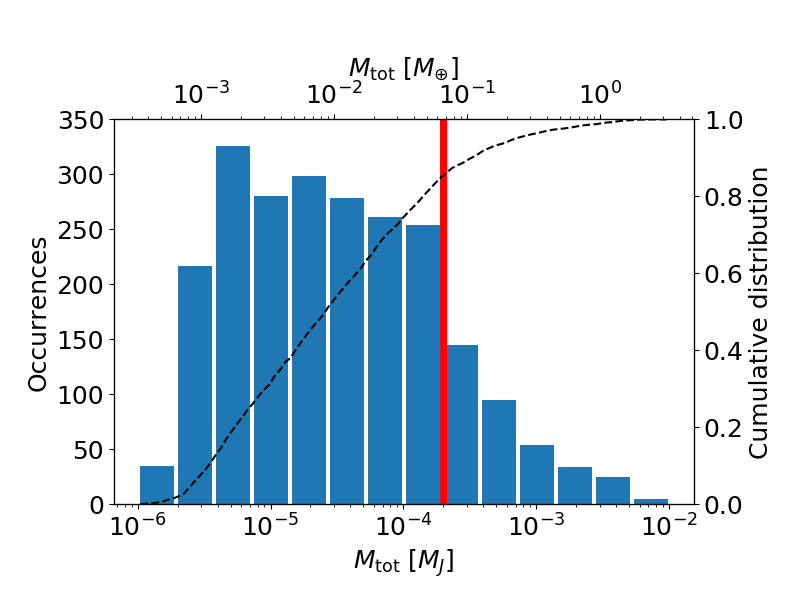}
\caption{Distribution of final total masses of satellites. The red line represent the total mass of Galilean moons, i.e. $2\times10^{-4}M_J = 0.06M_{\oplus}$. The black dashed line is the cumulative distribution. The results are taken from 2309 simulated systems.}\label{total_mass_distribution}
\end{figure}

\begin{figure}
\includegraphics[width=\columnwidth]{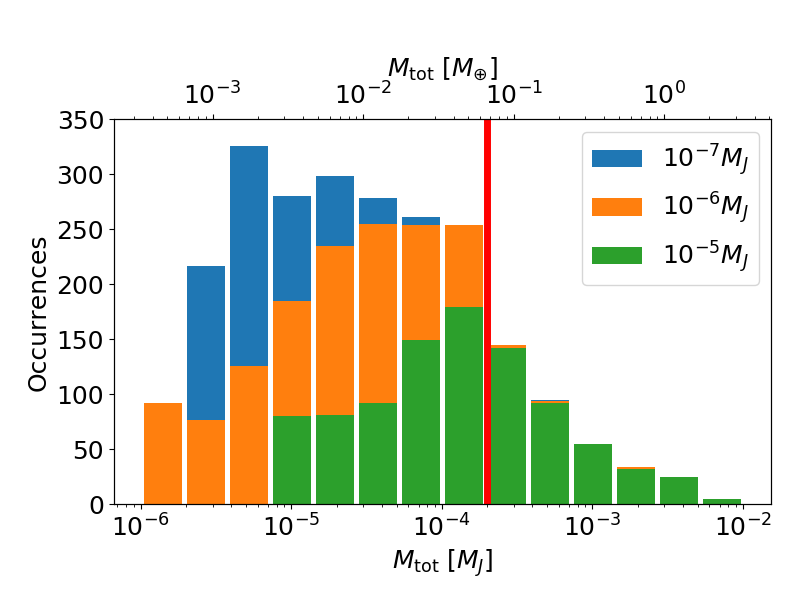}
\caption{Distribution of the final total moon masses including satellites only above a specific mass threshold ($10^{-7}M_J$, $10^{-6}M_J$, $10^{-5}M_J$). The red line represent the total mass of Galilean satellites, i.e. $2\times10^{-4}M_J = 0.06M_{\oplus}$. The black dashed line is the cumulative distribution.}\label{total_mass_distribution_threshold}
\end{figure}

During a single simulation, many satellites were engulfed by the central planet, collide with each other or get ejected from the system. First, we looked at the final mass of single survived satellites. From the histograms and the cumulative distribution in the first plot in Figure \ref{single_mass_distribution}, it is clear that small satellites are dominant and that the mass distribution is strongly peaked towards lower values. Nevertheless, most of this light satellites are just seeds that are produced throughout the simulation and have little effect on the real architecture of a satellite system. For example, the second plot of Figure \ref{single_mass_distribution} shows again the masses of single survived satellites, but the histograms are built weighting the satellites with their mass. In practice, the value of each column represents how much mass over the total mass of all the survived satellites is contained in that particular bin. In this case, we see that massive satellites are now dominant, meaning that, even though numerous, small satellitesimals are not important in determining and characterizing the structure of the survived systems.

Consequently, we identified the total mass of survived satellites as a more significant measure. Figure \ref{total_mass_distribution} shows that the low mass systems are more likely than higher mass systems. If we consider the Galilean system mass as a threshold (red line in Figure \ref{total_mass_distribution}), i.e. $2\times10^{-4}M_J = 0.06M_{\oplus}$, then $85\%$ of the produced systems are less massive than this threshold, while $15\%$ of the systems are more massive.

As all other statistical distributions in the population synthesis framework, these results highly depend on the choice of the values and distribution of initial parameters (dust-to-gas ratio and refilling timescale especially). 
Consequently, we present here more detailed analysis of the distribution of the results, while we are going to provide a study of the dependency between the chosen parameters and the results in Section \ref{Architecture of systems}.

For example, Figure \ref{total_mass_distribution_threshold} shows the total satellite mass distribution within the resulting moon-systems, but ignoring satellites less massive than the chosen thresholds. The figure shows that if the contribution of all the small satellitesimals, the shape of the distribution will change. In particular, the most massive systems do not vary a lot, because small satellites give only a small contribution, while the distribution changes quite much in the left tail. This is because low-mass satellites have a bigger impact on less massive systems. In particular, if only Galilean-like satellites ($M\ge10^{-5}M_J$) are considered, the peak of the (green) distribution is very close to the total Galilean mass ($2\times 10^{-4}M_J$).

This value ($\sim10^{-4}$) is also found to be the ratio between the mass of all the satellites and the mass of the planet in the case of Saturn, Uranus, and Neptune, even though the satellite system in the latter case is thought to have a completely different history, with Triton being captured rather than having formed in CPDs as in the other cases \citep{Goldreich89}. Since our model has been run only in the case of a forming Jupiter-like planet, we cannot derive strong conclusions, but these values suggest a correlation between the mass of the satellites and the mass of the planet. In the case of our model, that would mean that the solid content of a CPD (described by both the dust-to-gas ratio and the refilling timescale, as described in Section \ref{Architecture of systems}) would be more or less proportional to the mass of the planet. Nevertheless, the supply of solid materials to CPDs is not fully understood yet, and that is also why we need to rely on statistical analysis as in our Population Synthesis model.

\begin{figure}
\includegraphics[width=\columnwidth]{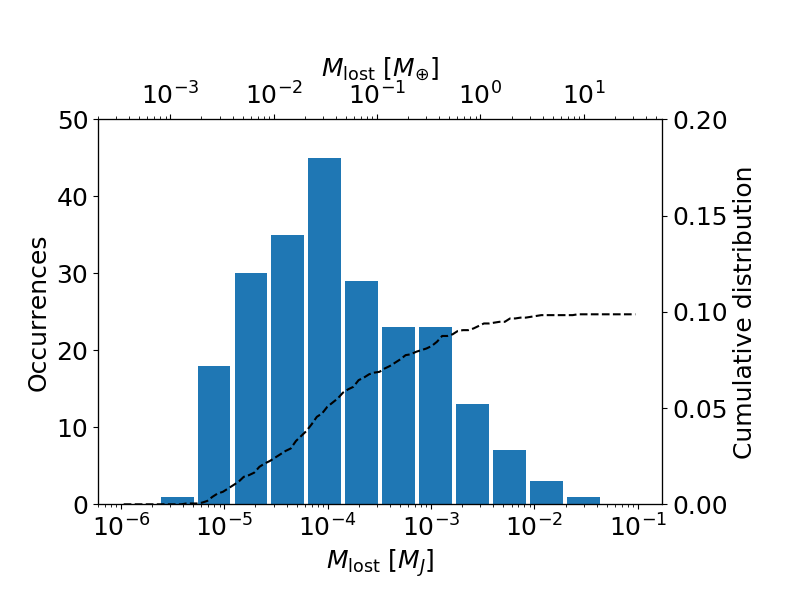}
\caption{Distribution of total lost satellite mass (engulfed by the planet). The black dashed line is the cumulative distribution. The results are taken from 2309 simulated systems.}\label{lost_mass_distribution}
\end{figure}

On Figure \ref{lost_mass_distribution} the total mass of satellites that were engulfed by the planet is shown. These moons contribute to the heavy element content of the gas giant planet envelope, and some probably sediment down to the core, possibly adding fuzziness to the core boundaries. This "metal-pollution" happens only in 10\% of the cases, i.e. in 10\% of the systems at least one satellite has been lost into the central planet. In 10\% of them, i.e. a total of 1\% of the total cases, the polluting solid-mass is comparable to $1M_{\oplus}$, about $0.003 M_J$, that is negligible compared to the $0.02-0.08M_J$ of solid material that Juno measured to be in the inner layers of Jupiter \citep{Wahl17}. From this we can conclude that solid accretion onto the planet is not as efficient as in our previous work \cite{Cilibrasi18}. The difference rises due to the absence of a dust trap in this work (Section \ref{Formation time}) and the 3D treatment of close-encounters and collisions (Section \ref{N-body interactions}) instead of 1D, that leads to less massive moons and, consequently, slower migration.

\subsection{Number and locations}\label{Number and positions}

\begin{figure}
\includegraphics[width=\columnwidth]{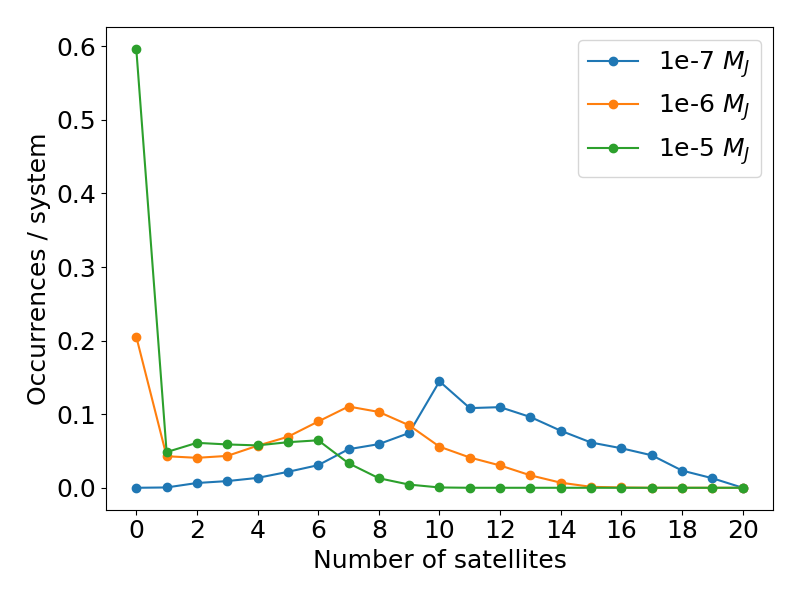}
\caption{Probability distributions of the number of survived satellites above a certain threshold mass ($10^{-7}, 10^{-6}, 10^{-5}M_J) $.}\label{number_distribution}
\end{figure}

The number of survived satellites and their semi-major axes were also examined. Figure \ref{number_distribution} shows the probability distributions of the number of survived satellites above certain threshold masses ($10^{-7}, 10^{-6}, 10^{-5}M_J$).
These thresholds are used to distinguish again satellite seeds that are always produced by the model, from full-grown moons, given the lack of a proper definition between the two mass categories. The distribution in Figure \ref{number_distribution} with $M_{\text{th}}=10^{-7}M_J$ (i.e. all satellites and satellite seeds) are for comparison purposes only. As the mass threshold is increased from the base value ($M_{\text{th}}=10^{-7}M_J$), there is a greater number of systems that cannot build any satellites above the threshold ($20\%$ for $M_{\text{th}}=10^{-6}M_J$, $60\%$ for $M_{\text{th}}=10^{-5}M_J$), and the shape of the distribution changes as well. For example, the case with the threshold of $M_{\text{th}}=10^{-5}M_J$ (i.e. counting only moons that are on the individual Galilean moon mass regime), about $18\%$ of the cases have 3, 4, or 5 satellites, in agreement with the Galilean system. This is trivial, since the same CPD can only produce smaller amount of heavier moons, or more numerous low-mass moons. 

\begin{figure}
\includegraphics[width=\columnwidth]{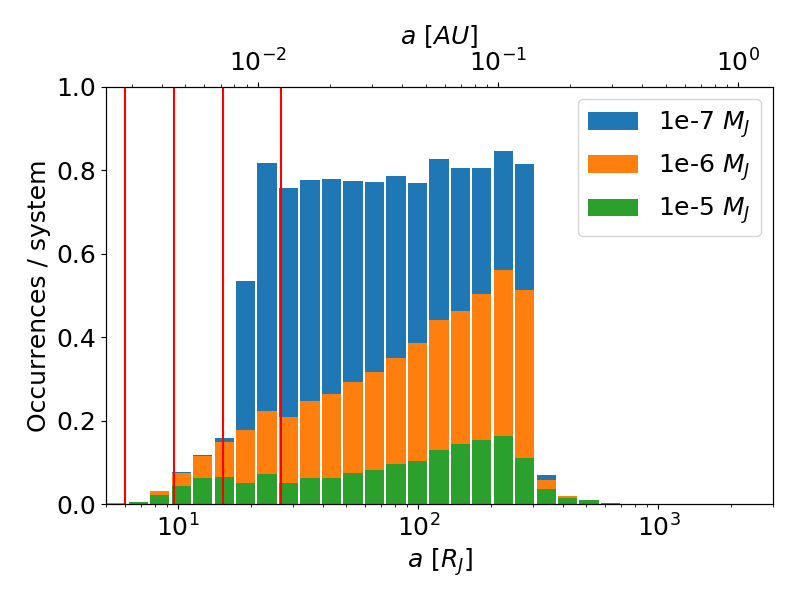}
\caption{Probability distributions of semi-major axes of satellites above a certain threshold mass ($10^{-7}, 10^{-6}, 10^{-5}M_J) $. The red lines are the current positions of Galilean satellites.}\label{semimajor_axis_distribution}
\end{figure}

In order to reproduce a Galilean-like system, it is also necessary to check if the satellites migrate from the formation location to the typical Galilean locations. The distribution of semi-major axes of satellites from our population (in Figure \ref{semimajor_axis_distribution}) shows that the distribution is mostly concentrated around the formation location because light satellites do not migrate much. When increasing the mass-threshold that we examine, the distribution tends to spread in the semi-major axis range a bit, especially towards the planet location. Satellites in the current Galilean-moons' locations increase from 15\% to 20\%. Nevertheless, the seeds that then reached the location of the Galilean moons were initially placed in the innermost part of the formation area.
Furthermore, in Figure \ref{semimajor_axis_distribution} there is a peak in the distribution in the outer part of the seed inserting region. This happens because migration is less effective in the outer part of the disc (the timescale is longer if computed as in Section \ref{Satellite-disc interactions}) and that causes satellites to migrate less when they form in that outer region. This is also the reason why the distribution spreads more towards the inner disc than towards the outer part when changing the mass threshold.

\begin{figure}
\includegraphics[width=\columnwidth]{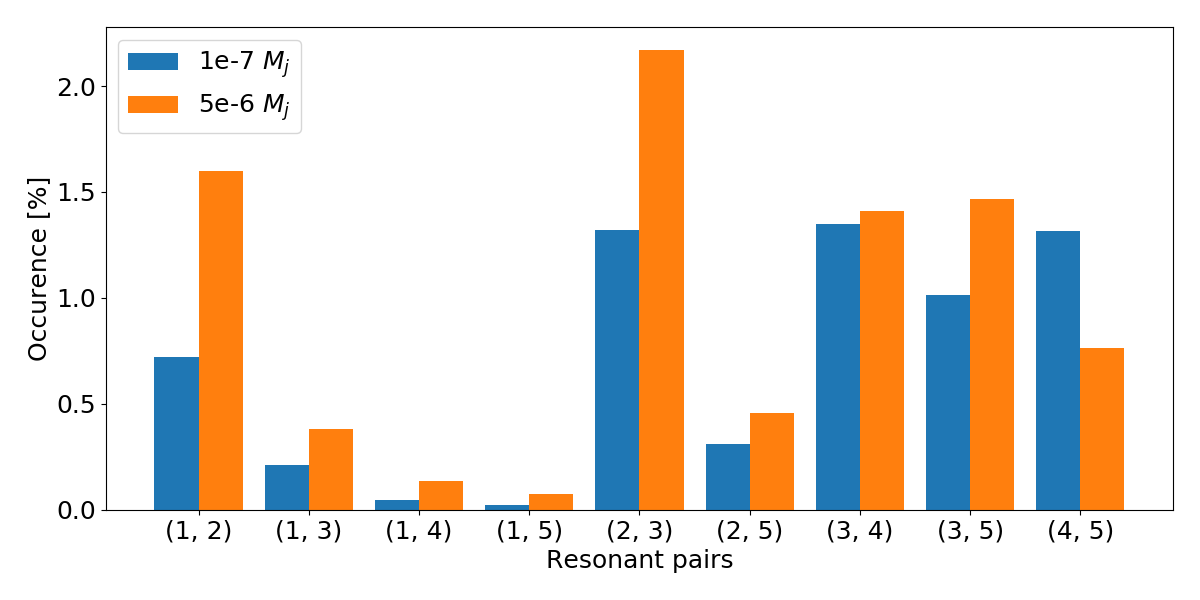}
\caption{Resonant pairs found in the simulations. When all full-grown satellites and satellite-seeds are included in the analysis, 6.3\% of them are found to be in resonance. When only massive satellites are considered, 8.5\% of them are in resonance, with the most common configuration being (1:2) and (2:3).}\label{resonance_plot}
\end{figure}

The semi-major axes of satellites are also linked to the resonances that may form in a satellite system, similarly to planetary systems. In fact, a mean motion resonance (MMR) occurs when orbiting particles or bodies exert a periodic and regular gravitational influence on each other, and that happens usually when their orbital periods, that is given by their semi-major axes, are related by a ratio of small integers. In this analysis, an approximate treatment has been implemented, and a pair of consecutive satellites $i$ and $j$ is considered as in a resonant configuration $p:q$, where $p,q\in N$, or in this case $p,q\in \{1,2,3,4,5\}$, when their period ratio is closer than $0.5\%$ to the ratio $p/q$ \citep{Hands14}.
We perform the analysis for all the satellites (about 24 thousand pairs) and for all satellites with $M>5\times M_{th}=10^{-6}M_J$ (about 5 thousand pairs), in order not to take into account the effect of small satellitesimals. As shown in Figure \ref{resonance_plot}, when all full-grown satellites and satellite-seeds are included in the examination, $6.3\%$ of the pairs were found to be resonant, while considering only the larger moons, there are $8.5\%$ of resonant pairs. In particular, $1:2$ and $2:3$ are the most common configurations.

\subsection{Architecture of systems}\label{Architecture of systems}

So far, distributions for moon-masses, -numbers and semi-major axes have been examined whether their values are close to the Galilean system. The previous analysis has two limitations: it depends on the distributions of initial and boundary parameters (dust-to-gas ratio, refilling timescale, initial number of satellite seeds), secondly, it does not quantitatively tell how similar a single produced system to the Galilean one. In order to overcome these limitations, it was defined how "close" a system is to the others or to the Galilean system, following the approach of \cite{Alibert19}. In particular, \cite{Alibert19} shows how a mathematical distance between the similarity of systems can be defined. First of all, each moon-system is assigned a 2D function depending on two variables, a mass $M$ and a semi-major axis $a$, defined as
\begin{equation}
    \psi_i(M,a) = \sum_{s \in i} f_{s}(M, a)
\end{equation}
with
\begin{equation}
    \begin{split}
        f_{s}(M, a) = &\exp\left[-\frac{1}{2}\left(\frac{\log M - \log M_{s}}{\sigma_M}\right)^2-\frac{1}{2}\left(\frac{\log a-\log a_{s}}{\sigma_a}\right)^2 \right]\cdot\\
        &\cdot F(M_s)
    \end{split}
\end{equation}
Here $f_s$ is a function calculated for each satellite ($s$) in the $i^{th}$ system, $M_s$ and $a_s$ are the mass and the semi-major axis of the satellite $s$, $\sigma_M$ and $\sigma_a$ are two arbitrary parameters that have to be tuned in the analysis, and $M$ and $a$ are the two aforementioned functional variables. The functions are defined over a reasonable domain, i.e. ($10^{-7}M_J\le M \le 10^{-2}M_J$, $1R_J\le a \le 1000R_J$), in order to cover all the possible combinations. 

In practice, the $\psi$'s are sums of 2D Gaussian profiles, one for each satellite in the system, with $\sigma_M$ and $\sigma_a$ defining the width of the Gaussians in the two directions. $F(M)$ is an arbitrary weight function that controls the height of the Gaussian peaks. In fact, the goal is to compare systems giving more importance to massive satellites, and that is why more massive satellites should have higher Gaussian profiles. We identified two possible weighting functions, depending on the desired application:
\begin{equation}
    \begin{split}
        &F_{\text{log}}(M_s) = \log_{10}(M_s/M_J) + 7\\
        &F_{\text{lin}}(M_s) = M_s / M_J
    \end{split}
\end{equation}
with the first function being the one used in this analysis.
Once every system has been associated to a function $\psi(M,a)$, a norm of these functions has been defined as
\begin{equation}
    ||\psi_i|| = \int\int|\psi_i(M,a)|^2 \rm{d}\log_{10}M\, \rm{d}\log_{10}a
\end{equation}
and as a consequence we have a distance, in arbitrary units,
\begin{equation}
    d_{ij} = ||\psi_i-\psi_j||
\end{equation}
This is proven to be an actual distance, fulfilling all the necessary conditions, as the functions are part of an $L^2$ space. Furthermore, we found convenient in the analysis to normalize the $\psi$'s in order to have $||\psi_i||=1$ $\forall i$.

\begin{figure}
\includegraphics[width=\columnwidth]{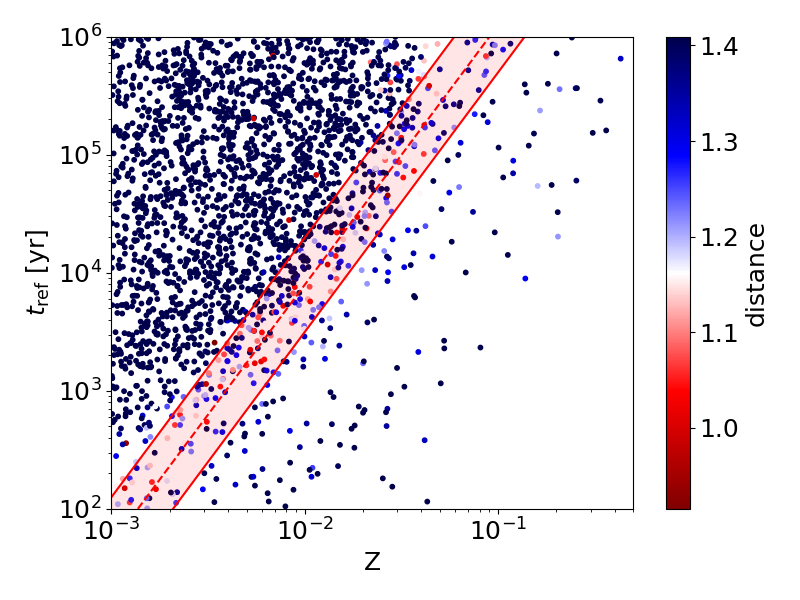}
\caption{Distance of all the systems from the Galilean satellites as a function of dust-to-gas ratio and refilling timescale. The unit of distances is arbitrary in this analysis framework. The red band highlights Galilean-like systems.}\label{distances_galileans1}
\end{figure}

\begin{figure}
\includegraphics[width=\columnwidth]{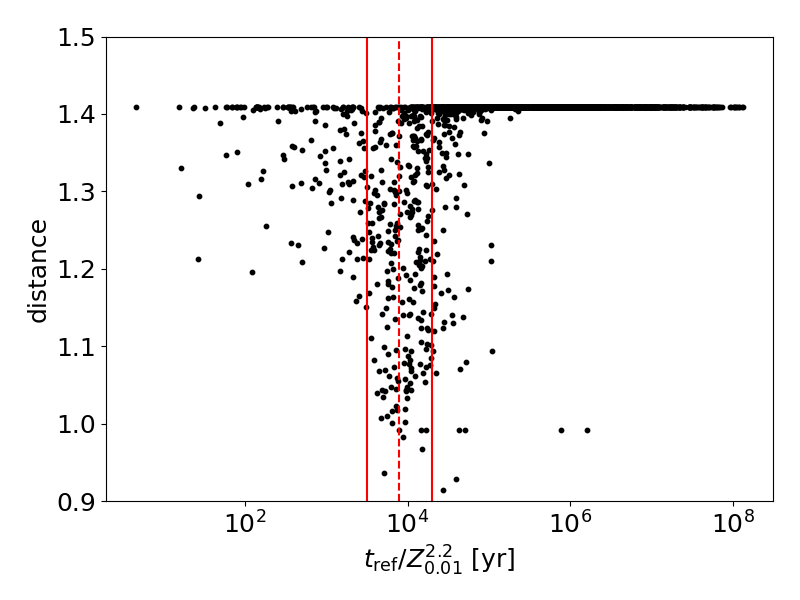}
\caption{Distance of all the systems from the Galilean satellites as a function of $\Gamma = t_{\text{ref}}/Z_{0.01}^{2.2}$. The unit of distances is arbitrary in this analysis framework}. The red lines highlight Galilean-like systems.\label{distances_galileans2}
\end{figure}

The first step of the comparison procedure is to check the distance of the individual moon-systems that came out from our population synthesis to the Galilean one. We defined a $\psi_G$ for the Jovian case and computed the distance $d_{iG}$ from all other systems $i$, using $\sigma_M = 0.1$ and $\sigma_a=0.1$. This means that two moons are considered very similar (i.e. their Gaussian profiles almost coincide and the difference gives a quasi-null integral) when they differ up to $1/10$ of order of magnitude in mass and semi-major axis.
All the distances $d_{iG}$ are shown as a function of dust-to-gas ratio and refilling timescale in Figure \ref{distances_galileans1}, as the initial number of seeds turns out not to have an effect, because the number of massive satellites is determined by the amount of solids in the CPD.
Given that normalized $\psi$'s can only have a distance between $0$ and $\sqrt{2}$, there are high distance points in the upper left corner and in the lower right in Figure \ref{distances_galileans1}, while there is a band, that is highlighted in red, where the distances from the Jovian system get smaller. In particular, measuring the slope of the band, we identify the Galilean-like "area" as the one approximately fulfilling
\begin{equation}
   8\times10^3 \text{yrs} \le\frac{t_{\text{ref}}}{Z_{0.01}^{2.2}} \le 5\times 10^4 \text{yrs}
\end{equation}
with $Z_{0.01}\equiv Z/0.01$ and with a peak around $\frac{t_{\text{ref}}}{Z_{0.01}^{2.2}}=2\times 10^4 \text{yrs}$. We can conclude then that the fundamental parameter, used to predict whether a system will produce Galilean-like satellite or not, is actually the ratio $\Gamma\equiv\frac{t_{\text{ref}}}{Z_{0.01}^{2.2}}$, as also shown in Figure \ref{distances_galileans2}. This relation can be verified visually from the plots, but it can also be derived more formally with the so-called Randomized Dependence Coefficient \citep{RDC}, as explained in Appendix \ref{Randomized Dependence Coefficient}.

In order to verify the aforementioned relation again, and in order to understand how really similar to Galilean satellites the produced systems are, we performed further analysis. In fact, being the distance in arbitrary unit by definition in this analysis framework, the previous analysis, shown in Figures \ref{distances_galileans1} and \ref{distances_galileans2}, tells how to produce the most Galilean-like systems in the model, in terms of their properties, but it is not clear how similar they are. It is necessary to understand whether, in those conditions, the distance from the Galilean system is similar (or even less) than the typical distance between the moon-systems produced by our population synthesis. Here, we are working in the infinite dimension space of $L^2$ functions in the chosen domain, hence it is not possible to easily visualize distances. Following the Machine Learning approach of \cite{Alibert19}, we also apply a T-distributed Stochastic Neighbor Embedding, or T-SNE, algorithm \citep{Vandermaaten08, Vandermaaten14} to our data. The T-SNE is a nonlinear dimensionality reduction technique meant for treating high-dimensional data for visualization in a low-dimensional space, typically a 2D arbitrary-unit space. In particular, minimizing a cost function, the algorithm links each high-dimensional object with a point in the 2D space so that similar objects are linked to close points and dissimilar objects are linked to distant points with higher probability. This procedure prevents from the typical mistakes of projections, where distant objects could be projected as being close to each other in a 2D plane.

\begin{figure}
\includegraphics[width=\columnwidth]{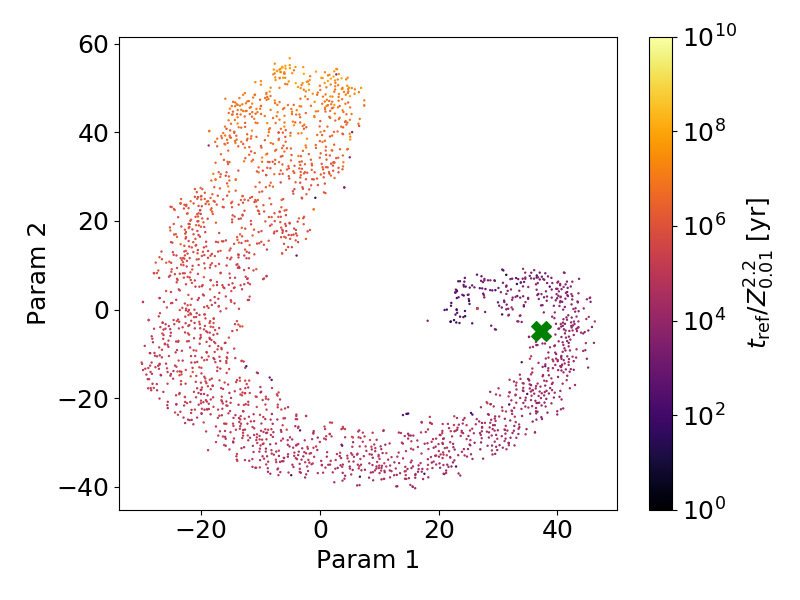}
\caption{Results of the t-SNE algorithm (see details in the text). The systems are reproduced in a 2D plot. The axes do not have any physical meaning, as the t-SNE approach is just a way to visualize high dimensional data in a 2D space, preserving local structures. Similar systems appear close and different systems appear far away. The green cross represents the Galilean system.}\label{TSNE_param}
\end{figure}

The t-SNE algorithm was applied to our data using the python package Scikit-Learn \citep{scikit-learn}, choosing a perplexity $p=50$, which is a parameter related to the number of nearest neighbors and balances the sensitivity of the algorithm to the local and global features of the data-set.
The result is shown in Figure \ref{TSNE_param}, in which the systems are reproduced in a 2D plot (the two axes do not have any physical meaning), so that similar systems appear close and different systems appear far away. The colour represents the value of $\Gamma = \frac{t_{\text{ref}}}{Z_{0.01}^{2.2}}$. First, the plot shows that the Galilean system (green cross) is well embedded in the cluster of systems. This means that the Jovian system is actually reproducible by our model in certain conditions. That would not be the case if the cross was out of the pipe-shaped cluster. Figure \ref{TSNE_param} shows also that the parameter $\Gamma$ is actually controlling the architecture of the final outcomes of the model. In fact, following the cluster from one end to the other, we have a continuous and gradual change of that parameter, with the Galilean system being close to $\Gamma = \frac{t_{\text{ref}}}{Z_{0.01}^{2.2}} = 2\times10^4 yrs$, as predicted before.

\subsection{Irregular satellites}\label{Irregular satellites}

\begin{figure}
\includegraphics[width=\columnwidth]{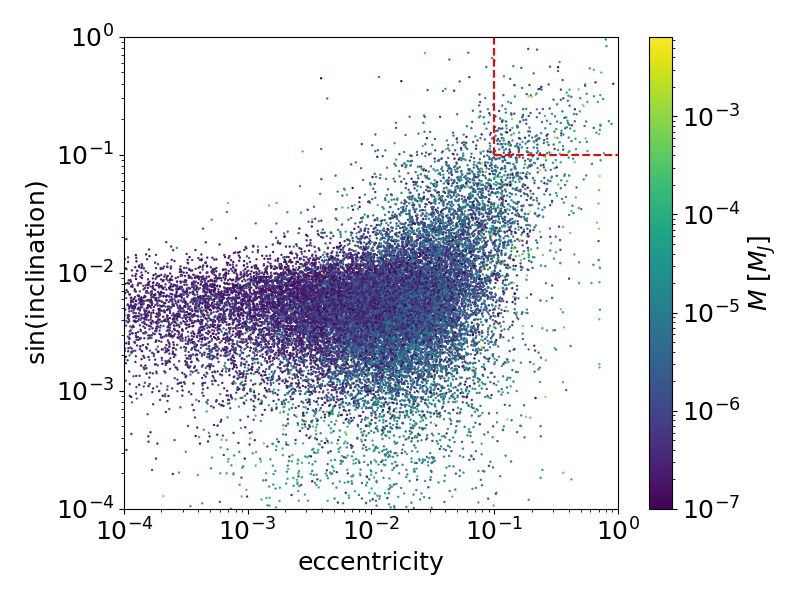}
\caption{Eccentricity and inclination of all satellites. The upper right includes what we define as irregular satellites.}\label{irregular_satellites1}
\end{figure}

Irregular satellites in the Solar System have large inclinations and eccentricities and are thought to be captured by the planets (instead of being formed around them) also because of their very low masses, of the order of asteroid masses. Our model could also produced some high inclination and high eccentricity moons, even though much more massive than asteroids ($\ge 10^{-7}M_J$), as dynamical interactions can toss them to highly inclined and elliptical orbits. 
We therefore checked the presence of such massive irregular satellites, that we do not see today around Jupiter, Saturn and Uranus (Neptune is believed to have had a more complex history in the satellite formation process, as shown by \citealt{Agnor06}). In our analysis, we consider a satellite as irregular when the eccentricity is $e\ge0.1$ and the inclination is $\sin(i)\ge0.1$, following the definition presented by \cite{Jewitt07}. With this definition, in the outcome of our population synthesis we found about $1\%$ irregular satellites (Figure \ref{irregular_satellites1}). In particular, our 2309 runs produced 26293 satellites, of which 262 are irregular. Among these, only one retrograde satellite was found, while the others were prograde.
In conclusion, our traditional moon-formation scenario within a CPD does not produce irregular massive satellites, in agreement with the configuration of moons of giant planets in the Solar System.

\subsection{Observability of exomoon-systems}\label{Observability of system}

Since our framework produces a population of satellites that formed in CPDs similar to Jupiter's disc, it is possible to check what fraction of the produced moons would be detectable with current and near-future instruments. Here we consider only the required sensitivies, not the observational time related feasibility. The planet in our model is at 5.2 AU from its Solar-equivalent star (Jupiter's distance), for which having 3 transits measured would require $\sim$ 36 years. 

For each system we computed which sensitivity is needed in order to have more than 50\% probability of observing that moon-system (assuming that the planet is transiting in front of the star from our point-of-view). The reason for this 50\% probability threshold is because the moons are not detectable during their entire orbits around the planet, due to the trivial geometrical effects. The sensitivity is considered as the instrumental capability to distinguish a transit curve from the noise. For example, an instrument is considered to have a sensitivity of $10^{-5}$ if it is able to spot a $10^{-5}$ single-transit dip in a stellar light curve. In this analysis, we identify how many moon-systems it is possible to detect given a sensitivity-limit, i.e. it is possible to infer a detection probability around a Jupiter-like planet as a function of the sensitivity. The details about the computation are in Appendix \ref{Observability model}.

\begin{figure}
\includegraphics[width=\columnwidth]{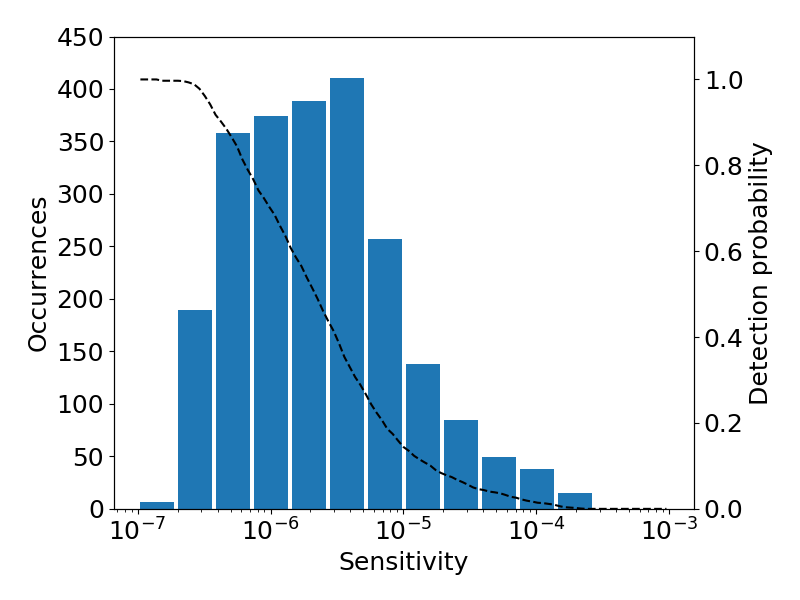}
\caption{Distribution of sensitivities needed to detect a moon-system. The black curve represents the fraction of systems that we would be able to detect with the given sensitivity. The analysis is on all 2309 systems.}\label{detection_histogram}
\end{figure}

Figure \ref{detection_histogram} shows the distribution of the minimal sensitivity that each system needs in order to be detected. The black curve on the figure represents the detection probability over the whole population with the given sensitivity. As expected, this is $1$ for very low sensitivities, because it allows to detect all the satellite-systems in our model, and it goes to $0$ for higher sensitivities. A sensitivity of $10^{-5}$ would allow to detect about 14\% of the systems produced within our model, while we find the median value at $2\times10^{-6}$ (a sensitivity of $10^{-6}$ would allow to detect more than 60\% of the moon systems).

\begin{figure}
\includegraphics[width=\columnwidth]{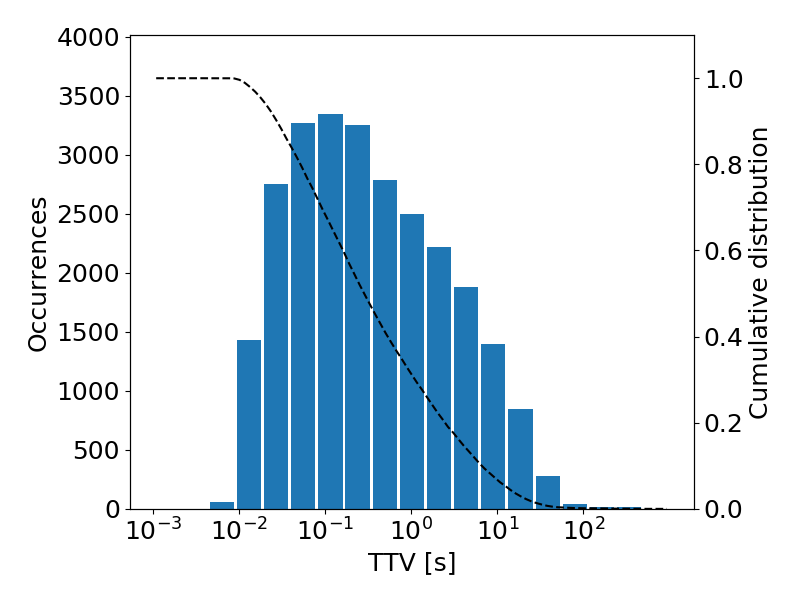}
\caption{Distribution of TTVs of satellites produced by the model on a Jupiter-like planet transit. The black curve represents the cumulative distribution.}\label{detection_TTV}
\end{figure}

In order to have a more general insight into the detectability of moons, we also investigate the magnitude of the TTV (Transit Timing Variation) that the satellites produced by our model would have on a Jupiter-like planet transits. In fact, transits alone could not distinguish between a moon or a planet transit. On the other hand, timing techniques are able to disentangle the two types of events. In particular, with TTV alone is not possible to detect all the physical features of exomoons, as it depends on the product of their mass and their semi-major axis, but a TDV (Transit Duration Variation) is needed to get this information \citep{Kipping09}.
The TTV effect is calculated assuming zero eccentricity, i.e.
\begin{equation}\label{TTV_equation}
    \delta_{TTV}\simeq 0.1 \frac{a_s}{a_p} \frac{M_s}{M_p+M_s} P_p
\end{equation}
where $a_s$ is the semi-major axis of the satellite, $a_p$ the one of the planet, $M_s$ and $M_p$ the masses of the satellite and the planet, $P_p$ the orbital period of the planet \citep{Kipping09}.

The TTVs calculated for our population are shown in Figure \ref{detection_TTV}. The distribution spreads from 0.01 seconds to over 100 seconds, has a mean of 2.56 sec and a median of 0.26 sec. Furthermore, as the cumulative distribution in Figure \ref{detection_TTV} shows, about 6.2\% of the satellites would produce a TTV greater than 10 sec. For comparison, \cite{Kipping09, Rodenbeck20} showed that a dozen of seconds would be a reasonable resolution for this kind of measurements.

As mentioned at the beginning of this paragraph, these observability tests have been done assuming a Jupiter-like planet ($1M_J$ at $5.2$ AU from a Sun-like star). However, most of, but not all, the giant planets we know today are found much closer to their parent stars. In those cases, both this analysis and the whole model presented in this work would not be reasonable any more. First of all, the structure of the CPD would be very different. In the inner part of a planetary system, Hill spheres are smaller and that means that also CPDs must be smaller and more compact. At the same time, the higher temperatures would favor the formation of a circum-planetary envelope more than a disc \citep{Szulagyi16}, and that would change the whole dynamics of solids and satellites. Furthermore, if we consider planets forming in the outer part of the disc and then migrating inwards, \cite{Bolmont19} showed that survival of satellites should be rare. and satellites should more likely be captured afterwards than form in situ.


\section{Discussion and Caveats}

This Section describes the major caveats and biases of our framework, identifying their effects on the results and comparing them to previous literature. In particular, we discuss the choices of the migration formulas and the chosen solid density profile, we compare our model with a simplified 1D model, we analyse the differences between a 1D and a 2D disc model, and we identify future developments of the model.

\subsection{Migration prescriptions}\label{Migration formulas}

As explained in Section \ref{Satellite-disc interactions}, the parameter for type I migration $b$ has been taken from \cite{Jimenez17}, because it was derived from 3D non-isothermal simulations, and can be applied to low- and intermediate-mass planets in optically thick discs. On the other hand, other migration formulae can be found in the literature (\citealt{Tanaka02}, \citealt{Dangelo10}, \citealt{Paardekooper10, Paardekooper11}), with the latter being the one used in our previous work, in \cite{Cilibrasi18}. The Paardekooper-formula was derived from simulations with low-mass planets embedded in 2D non-isothermal discs, and here we compare it with the one by \cite{Jimenez17}, showing that they give very similar migration rates in our domain. Therefore the differences between this work and \cite{Cilibrasi18} are not due to the different migration timescales.

\begin{figure}
\includegraphics[width=\columnwidth]{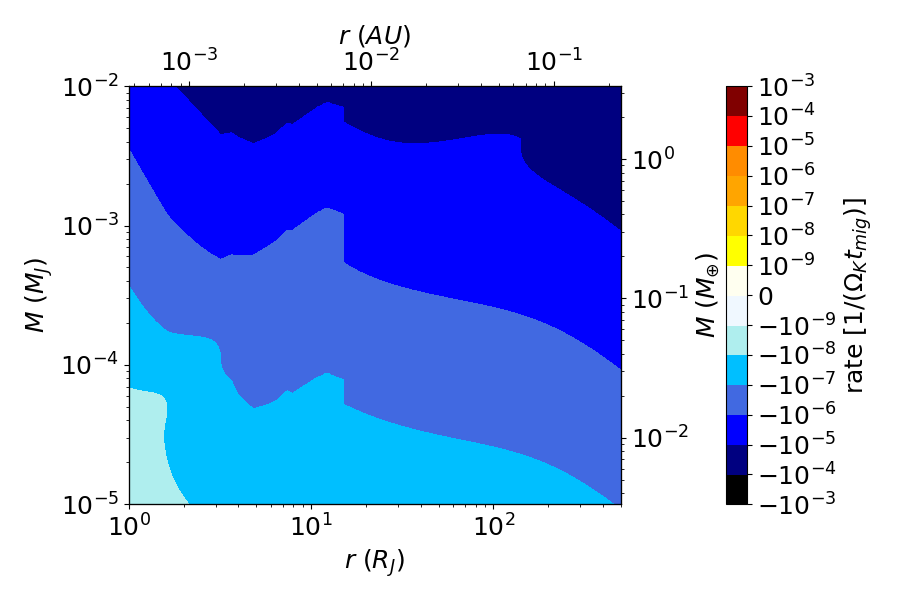}
\caption{Migration rates as calculated with the formula from \protect\cite{Jimenez17}. The rate is the inverse of the migration timescale multiplied by the orbital frequency, i.e. it is the orbital timescale divided by the migration timescale.}\label{migration_jimenez}
\end{figure}

\begin{figure}
\includegraphics[width=\columnwidth]{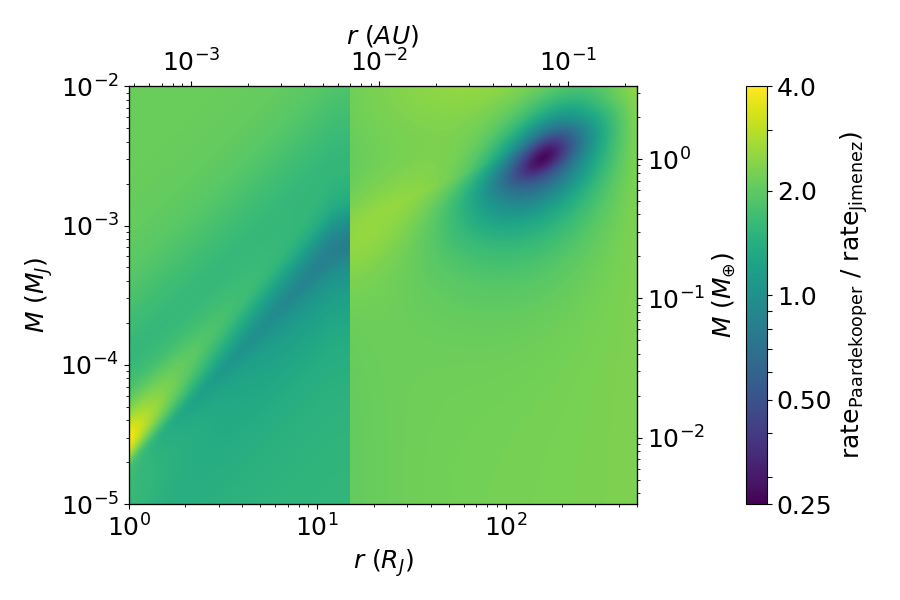}
\caption{Migration rate comparison between formulae from \protect\cite{Paardekooper10, Paardekooper11} and \protect\cite{Jimenez17}. The rate is again the inverse of the migration timescale multiplied by the orbital frequency, i.e. it is the orbital timescale divided by the migration timescale.}\label{migration_comparison}
\end{figure}

The migration rate, as calculated following \cite{Jimenez17}, is shown in Figure \ref{migration_jimenez} as a function of the satellite semi-major axis and mass. The plot shows that migration is always inward (always negative) and it is always very slow. In fact, the ratio between the orbital timescale and the migration timescale is between $10^{-9}$ and $10^{-5}$. Figure \ref{migration_comparison} shows a comparison between the migration rates as calculated with \cite{Jimenez17} and \cite{Paardekooper10, Paardekooper11}. In this case we see that the ratio of the two rates is always around 1, with a maximum spreading between 0.25 and 4, in very localized areas in the parameter space. This means that the two rates are always comparable well within one order of magnitude and we do not lead to significant differences for our population synthesis outcome. The plots also show the transition between the two different slopes of the temperature profile. In fact, the sharp transition occuring at around $15 R_J$ in Figure \ref{migration_jimenez} and Figure \ref{migration_comparison} is due to the temperature maximum (and change of slopes) at the same distance in Figure \ref{temperature_profile}.

\subsection{Formation time and dust-trap effect}\label{Formation time}

\begin{figure}
\includegraphics[width=\columnwidth]{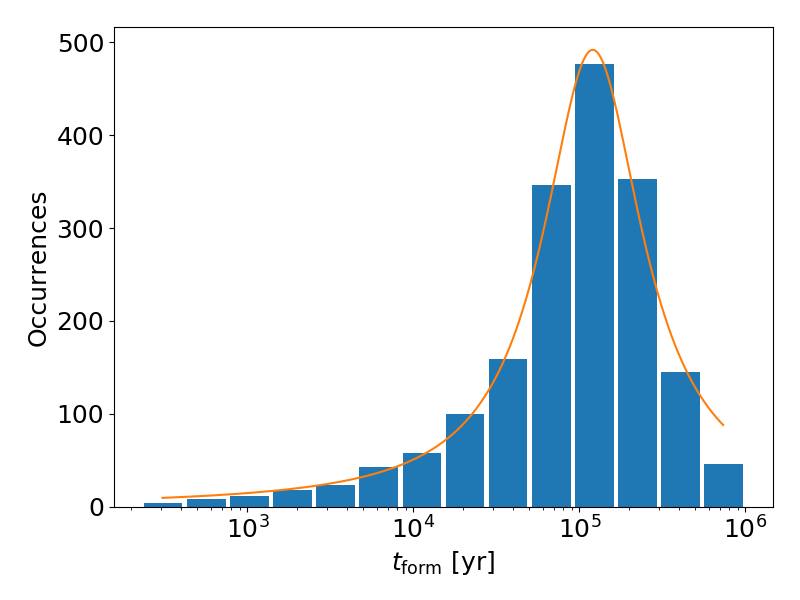}
\caption{Formation timescale for massive satellites, i.e. the time required to reach a mass of Europa. The orange line is a Cauchy distribution fitted to the histogram.}\label{formation_timescale}
\end{figure}

In the resulting moon population of our model, the formation timescales of satellites were also examined. Because the different massive moons grown on different timescales, we chose to look into how much time it takes to grow to Europa's mass, which is the lightest Galilean moon ($2.53\times 10^{-5}M_J=8\times10^{-3}M_{\oplus}$). The formation timescale histogram in Figure \ref{formation_timescale} shows 1792 satellites, that have reached this threshold mass. The histogram shows a one-peak distribution with a long tail on the left, toward smaller timescales, and a shorter tail on the right, given by the few cases in which satellites grow due to collision until the end of the simulation (after $10^6$ years). Furthermore, the histogram can be fitted with a Cauchy distribution $\propto \left\{1 + \left( \frac{\rm{log}_{10}t_{\rm{form}}-\mu}{\sigma} \right)^2 \right\}^{-1}$ with $\mu = 5.08 \pm 0.01$ and $\sigma = 0.37 \pm 0.02$, being the peak of the distribution at about $10^5$ years. In particular, about $53\%$ of the satellites take more than $10^5$ years to reach the Europa's mass. If we restrict our analysis to satellites with masses between Europa ($\sim2.4\times10^{-5}M_J$) and Ganymede ($\sim7.4\times10^{-5}M_J$), the shape of the distribution does not change substantially, but satellites form more slowly on average, with about $75\%$ of the moons growing in more than $10^5$ years.

Furthermore, the formation timescale is the only quantity that has been found to be dependent on the dispersion timescale by \cite{Cilibrasi18}. Nevertheless, this effect was mostly due to sequential formation, a process that has not been detected with the model presented in this work. This confirms again the choice we made, i.e. $t_{\rm{disp}} = 10^5$ yr.

Hundred-thousand years is a special threshold for satellite formation timescales, given the internal structure of the Galilean moons. In particular, we know that the first three of them (Io, Europa, Ganymede) are fully differentiated, while Callisto is only partially differentiated \citep{Anderson01}. These different features are thought to be linked with the formation timescale. A fast formation and accretion can cause a satellite to melt, because of the high rate of energy absorption, allowing heavier elements to sink and form layers. On the other hand, a slow formation process would not cause melting, preventing the moon from differentiation. The threshold accretion time that divides the two scenarios is thought to be $10^5$ years in some models \citep{Stevenson86, Canup02}.

Compared to our previous work, \cite{Cilibrasi18}, the peak of the distribution in Figure \ref{formation_timescale} is considerably shifted to the right, showing much slower formation timescales this time. This is mainly due to one difference in the two models, i.e. the solid distribution in the CPD (other minor differences, such as the 3D vs 1D dynamical integration, turns out to have a minor effect and are investigated in Section \ref{1D vs 3D}). While the model presented in this paper consider the solid distribution to have the same shape and slope of the gas distribution, the model in \cite{Cilibrasi18} used a dust distribution derived from a 1D dust evolution model by \cite{Drazkowska18}, that shows a peak in the distribution at about $85R_J$ from the planet. This location is a dust trap and was considered to be the main location where seeds formed within the CPD. Here, the moonlets acquired most of their mass, which happened very quickly, given the high concentration of dust. In this paper, the smoother solid density profile allows seeds to form everywhere in the disc, but that leads to slower growth rate. Consequently, this difference also explains why there is no frequent \textit{sequential formation} in this work. Slowly growing satellites do not experience fast Type I migration, and hence less of them gets engulfed by the planet. 

\subsection{1D vs 3D}\label{1D vs 3D}

In order to understand the importance of using an N-body integrator in our model, we compared our results with runs performed in a fully 1D population synthesis model. In this test, the disc evolution was treated exactly the same way, while the satellite dynamics was treated as in \cite{Cilibrasi18}, assuming circular orbits, migration rates by \cite{Jimenez17} and resonant trapping as described by \cite{Ida10}. Collisions were triggered when two satellites approached one another at a distance smaller than the sum of their respective Hill-radii. Because the computational time is orders of magnitude lower than of the 3D runs, we were able to perform a 1D simulation starting from the same initial conditions and parameters for each of the 3D simulations.

\begin{figure}
\includegraphics[width=\columnwidth]{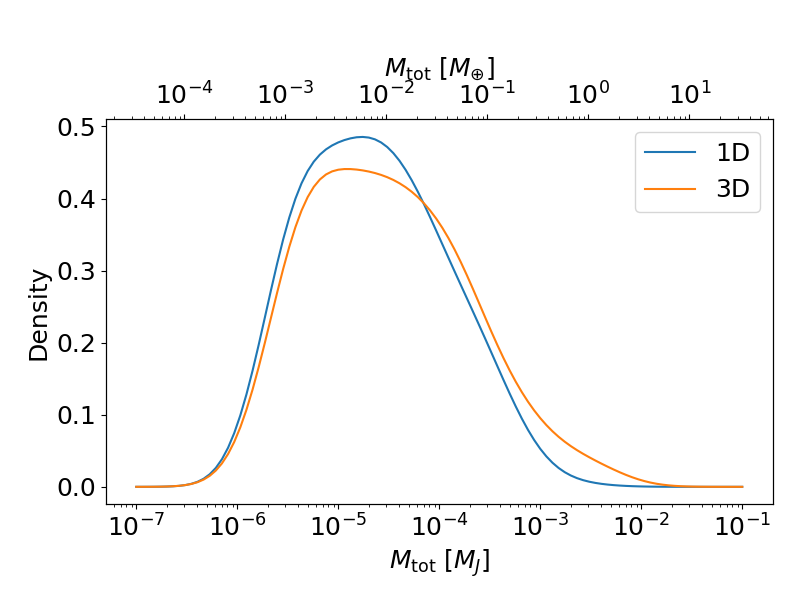}
\caption{Distribution of the total mass of survived satellites (at the end of the simulations) in each system in the 1D and 3D model. The distribution is normalized so that the integral is 1, as we consider the same number of systems in the two cases.}\label{total_mass_distribution_1D_3D}
\end{figure}

\begin{figure}
\includegraphics[width=\columnwidth]{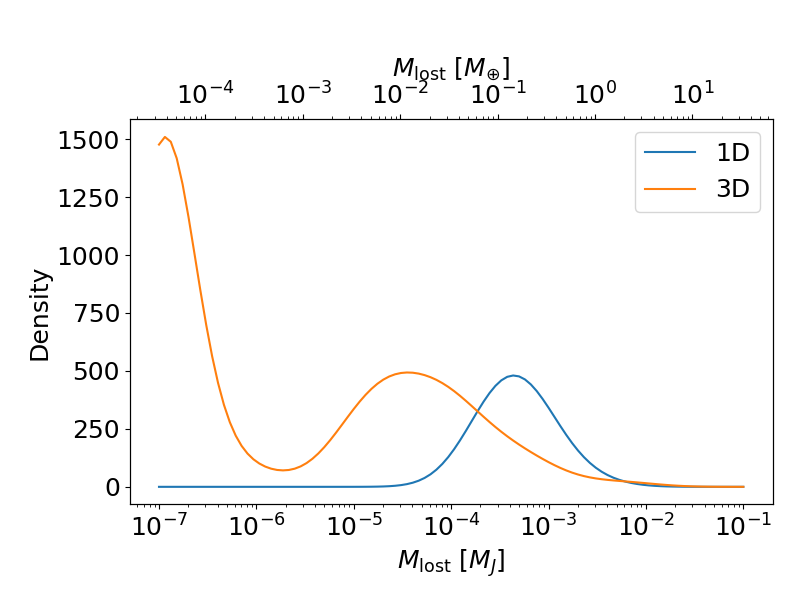}
\caption{Distribution of the mass of all lost satellites that are engulfed by the planet. The distributions are normalized such that the integral equals the total number of lost satellites, in order to highlight that lost satellites are more numerous in the 3D model, and the shapes of the distributions are quite different.}\label{single_lost_mass_distribution_1D_3D}
\end{figure}

We compared the distributions of the total satellite mass in each system (Figure \ref{total_mass_distribution_1D_3D}). The difference is that the satellites can grow to larger masses in the 3D model relative to the 1D model. The reason for this is mainly how collisions are treated and can be counter-intuitive at first. In 1D the collisions are trivial, while in 3D the collisional cross-sections are significantly smaller. In the first case, this accelerates the mass growth of moonlets, which then increases their Hill-radius, leading to a further increase in the collision rate and mass. This can be seen as a rapid "runaway" accretion process. However, because of the larger masses, satellites also migrate faster towards the planet, eventually getting engulfed by the gas-giant, then not surviving. On the other hand, in the 3D model, this runaway growth is not occurring and satellites can get massive without igniting the rapid accretion, and, eventually, surviving. At the same time, in the 3D model, collisions can scatter small satelletesimals towards the planet, that would not be affected by inward migration given their low masses. Scattering is completely absent in the 1D model due to the dimensional limitation.

Figure \ref{single_lost_mass_distribution_1D_3D} shows the mass distribution of moons that are lost into the central planet. This figure again demonstrates the different outcomes of the 3D and 1D models, similarly to the explanation above. The curve of the 3D model shows two peaks, one for small moons, which are delivered to the centre by scattering, and one for larger masses, which
are the result of inward migration. On the other hand, the curve for the 1D model shows only one peak, which is caused by migration only, since there is no inward scattering in 1D. 

\begin{figure}
\includegraphics[width=\columnwidth]{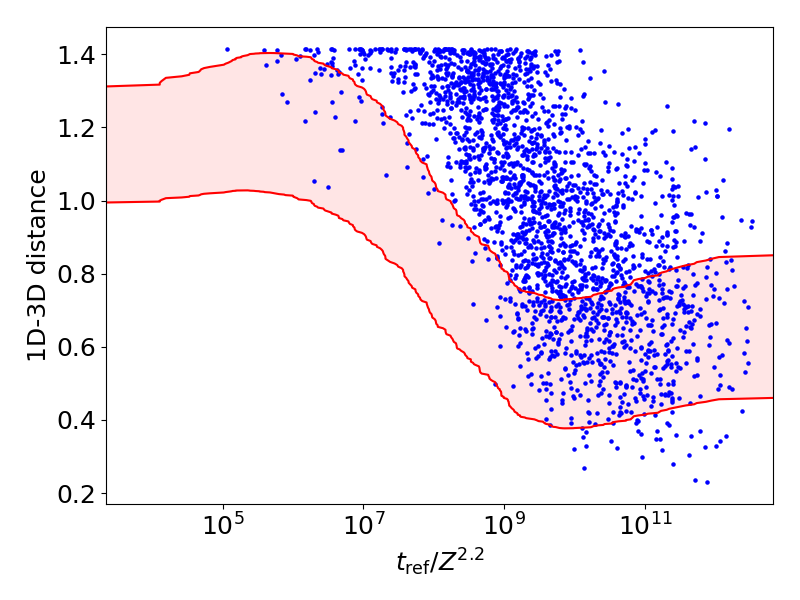}
\caption{The blue dots represent the distance between the results obtained with 1D and 3D moon dynamics in each system, while the red area is the distance dispersion range of systems generated by the same initial parameters in the 1D model
}\label{1D_3D_distance}
\end{figure}

In order to understand more qualitatively the difference between the 1D and 3D treatments, the same machine learning framework was used as described in Section \ref{Architecture of systems}. Thanks to the fast running time of the 1D simulations, we performed 20 simulations for 300 different combinations of $Z$, $t_{\rm{ref}}$ and $N_{\rm{init}}$, each time starting from an initial random distribution of seeds. This way it is possible to estimate the spreading of systems simulated with the same parameters, i.e. the dispersion of results that is naturally caused by the randomness of the code (initial seed positions), and compare it to the distance between the 1D and the 3D model. Using $\Gamma=t_{\rm{ref}}/Z_{0.01}^{2.2}$ as the reference parameter, we show this comparison in Figure \ref{1D_3D_distance}. The distance between 1D and 3D results (blue dots) is compared with the typical dispersion range of distances between systems generated from the same parameters (red range), making clear that the two values are comparable only in the range where $t_{\rm{ref}}/Z_{0.01}^{2.2}$ is high, i.e. where satellites grow and migrate less, while they are well-distinguished for lower values. This means that the 1D and the 3D models produce significant differences.

\subsection{1D vs 2D Accretion}\label{1D vs 2D Accretion}

\begin{figure}
\includegraphics[width=\columnwidth]{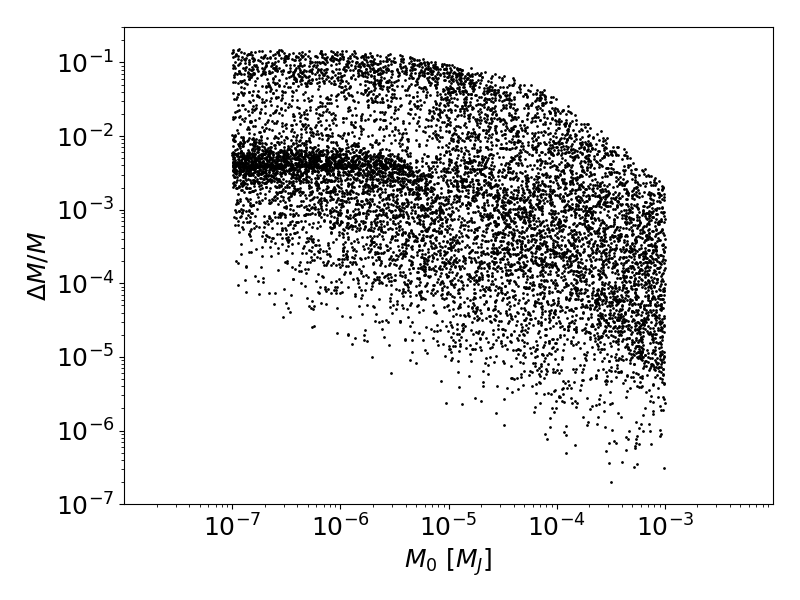}
\includegraphics[width=\columnwidth]{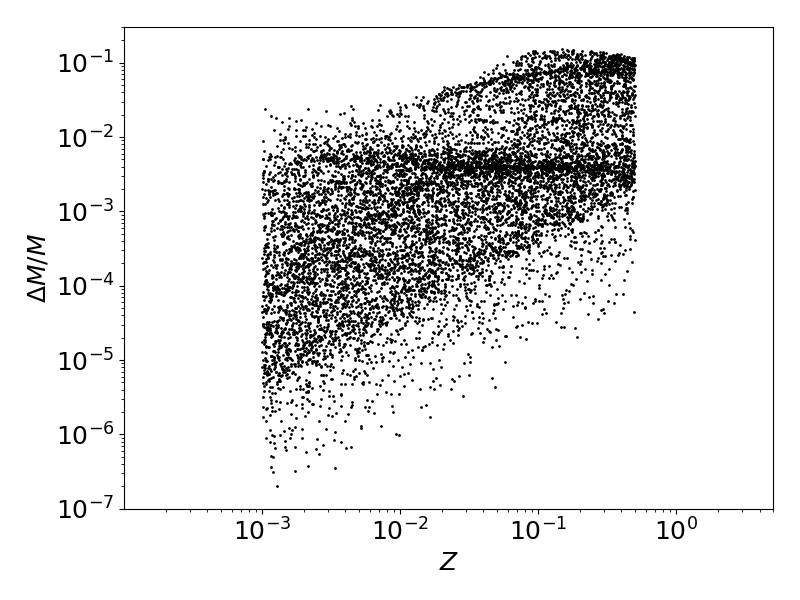}
\caption{Ratio between masses in simulations with 1D and 2D moon dynamics as a function of the initial embryo mass (top panel) and the dust-to-gas ratio of the disc (bottom panel).}\label{mass_ratio_scatter}
\end{figure}

\begin{figure}
\includegraphics[width=\columnwidth]{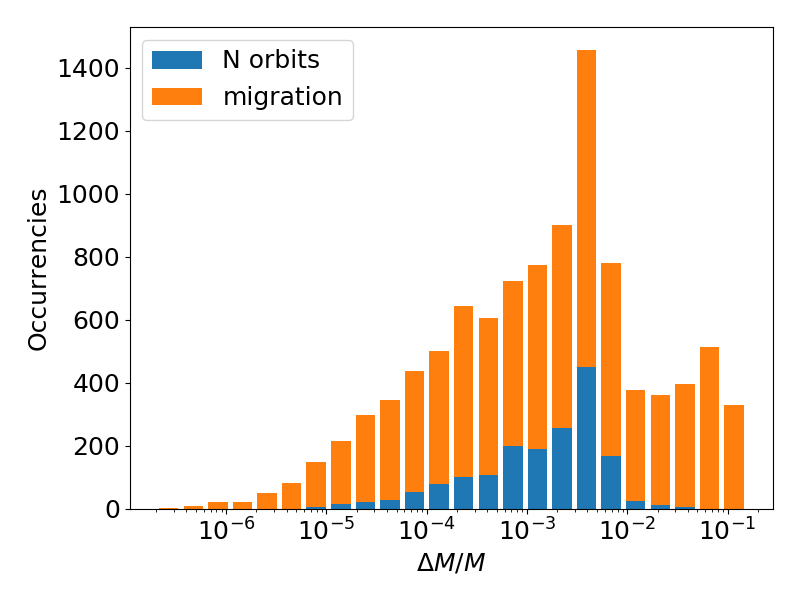}
\caption{Distribution of the relative difference between 1D- and 2D-calculated moon-masses. The peak is at about $0.2-0.5\%$ while the maximum difference is at about $15\%$. The two colours identify satellites that stop the accretion because of migration and the ones that stop because of reaching N=100'000 orbits (i.e. end of the simulation).}\label{mass_ratio_histogram}
\end{figure}

We also tested what differences it would make if the CPD was treated in 2D, instead of 1D. Since our disc model is coming from a 3D thermo-hydrodynamical simulation, it is possible to consider the midplane of this disc, and evolve it keeping its 2D structure (e.g. the spiral arms present in the disc).

We verified in the 2D case too, that migration timescales are much longer than orbital timescales. This means that assuming a 1D profile is giving just a negligible correction to migration and damping calculations compared to a 2D model, since the orbital motion of the Keplerian satellites through the sub-Keplerian gas is already averaging all disc quantities azimuthally.

On the other hand, solid accretion is different in 2D and 1D. In fact the satellites and the solids are in principle orbiting with the same Keplerian speed and do not necessarily show the same azimuthal averaging process. This means that a moonlet does not have access to all the solids in its orbit at once, as in the 1D model, but only to the amount in its 2D feeding zone ($R_{\text{feed}}=2.3\,R_{\text{Hill}}$; \citealt{Greenberg91}). Nevertheless, one can expect the differential velocity of all the solids in the feeding zone to produce some azimuthal average as well. In order to understand whether or not this average process actually happens, we need to calculate how different accretion is in the two scenarios (1D versus 2D). The process is not linear, since a satellite would accrete more if it is more massive (since the Hill radius would be larger, hence its feeding zone as well). To test the difference that this makes, we set up 10'000 pairs of simulations. Each pair had the same satellite with a mass randomly chosen between $10^{-7}M_J$ and $10^{-3}M_J$, a semi-major axis between $10R_J$ and $300R_J$, embedded in a disc with a dust-to-gas ratio between $0.001$ and $0.5$ and refilling timescale between $10^2$ yrs and $10^6$ yrs. In each of the two simulations in the pair, the satellite was let free to accrete mass from the solids of the disc, until one of the two following conditions happened:
\begin{itemize}
    \item the time in the simulation had reached 100'000 orbits at the location of the satellite
    \item the satellite had migrated more than $0.1\%$ of its semi-major axis because of migration, calculated as Section \ref{Satellite-disc interactions}.
\end{itemize}
In the pairs, one simulation was performed with a 2D orbiting (Keplerian) solid density grid and the other simulation was performed with a 1D grid, with the results being compared afterwards. In the analysis, the initial mass of the embryos and the dust-to-gas ratio of the disc turned out to have the biggest influence, while the semi-major axis and the refilling timescale did not have a fundamental effect.

The dependence on the initial mass of the satellite seeds and the dust-to-gas ratio is shown in Figure \ref{mass_ratio_scatter}. As expected, higher dust-to-gas ratios produce higher differences in the final mass ratio: the more massive moons accrete faster and create a bigger discrepancy. On the other hand, embryos have a different evolution in the very first phases, when the mass is still small. Once satellites become more massive, then this effect is much smaller. The plot shows that the difference between the 1D moon masses (being the larger) and the 2D ones is never more than 15\%. In the mass ratio histogram in Figure \ref{mass_ratio_histogram} between the 1D and 2D cases, we see that the difference is more than $1\%$ in $19\%$ of the cases, while it is between $10-15\%$ in $2\%$ of the cases. This means that having a 1D solid profile instead of a 2D would produce a significant difference in the accretion only in $2\%$ of the cases. Once the small satellites reach higher masses and start to migrate through the disc, then the difference becomes even smaller.

On Figure \ref{mass_ratio_histogram} the high peak in the mass ratio histogram (that is linked to the over-density areas in Figure \ref{mass_ratio_scatter}) is due to a bias in the set-up of the simulations. Even with high dust-to-gas ratios, in cases when refilling is slow, the disc is able to provide a maximum of about $10^{-4}M_J$ of solid mass to the satellite before migration starts. Satellites often reach that maximum mass and either they stop their accretion (small initial mass), keeping the same $\Delta M/M$ until the end of the simulation, or the simulation stops because of the migration timescale constraint (big initial mass). This "saturation" value of $\Delta M/M$ turns out to be $5\times10^{-3}$ and it is mostly due to the exactly co-rotating solids.

\subsection{Biases and Future Developments}\label{Biases}

\begin{figure}
\includegraphics[width=\columnwidth]{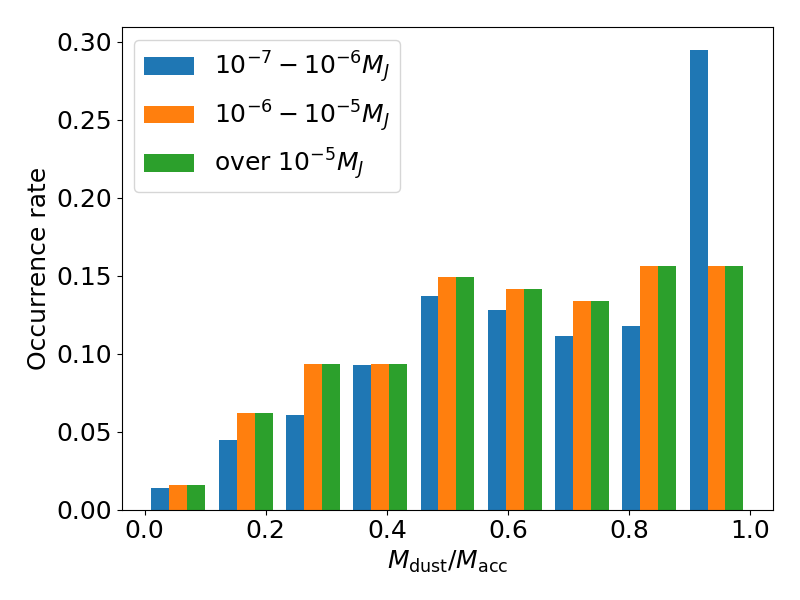}
\caption{The histogram shows the fraction of the total mass ($M_{\rm{tot}}$) gained by accretion ($M_{\rm{dust}}$). Most of the mass of moons gained by accretion from the solids disc, as opposed to collisions. The different colors represents the different moon-mass classes in the model. Small satellites (blue bars) grow mainly because of accretion (as shown by the peak around $0.9$), while bigger satellites experience collisions more often, even though most of their mass still comes from accretion.}\label{dust_accretion}
\end{figure}

In Section \ref{Satellite-disc interactions}, the accretion model is a relatively simple one, but accretion itself indeed is a very important process, that affects the outcome. This can be seen from Figure \ref{dust_accretion}, which represents the fraction of mass accreted from the disc of solids over the total mass gained. It is clear, that most of the proto-satellites grow because of accretion, less because of collisions with other moonlets. The amount of solids accreted is of course affected by the density profile evolution, the disc dispersion time and the refilling mechanism (i.e. how much solids are getting into the circumplanetary disc from the meridional circulation). Instead of our oligarchic growth method, other accretion schemes could have been implemented too. For example, using pebble accretion \citep{Ormel10, Lambrechts12}, which was already applied in the satellite-formation models of \cite{Shibaike19, Ronnet20}. 

One of the main missing element, is the lack of dust-gas hydrodynamical simulations of circumplanetary discs, that can inform us about the global and local dust-to-gas ratios and the solids flows based on particle size. Furthermore, our knowledge on the dust coagulation in the circumplanetary disc is also very limited \citep{Drazkowska18}. These effects of course very much affect any moon-formation model, since the satellites are forming from solids.

Because the base hydrodynamical simulations were boundary layer accretion simulations, there is no cavity between the planet and the circumplanetary disc in our model. Such a cavity would affect the formation of resonant chains \citep{Ogihara09,Ogihara12,Sasaki10,Fujii17}.

\section{Conclusions}

In this work, we investigated the formation and the evolution of satellites in a circumplanetary disc around a Jupiter-like planet, implementing a semi-analytical model for the disc evolution, and an N-body integrator for satellite-dynamics. We used a population synthesis approach, then statistically analysed the outcome. 

In particular, the circumplanetary disc density and temperature profiles were taken from radiative 3D hydrodynamical simulations \citep{Szulagyi17gap} and adapted into 1D and 2D grids. The solid profile was taken as a fixed fraction (dust-to-gas ratio) to the gas density, and the density and temperature profiles in the model evolve following an analytical (exponential) evolution. The proto-satellites formed and evolved interacting with the disc and consequently experiencing migration, eccentricity and inclination damping \citep{Cresswell08}, and interacting with each other (using an N-body integrator with individual time-steps \citep{Saha94} and a close-encounter treatment \citep{Chambers99}). Satellites also accreted mass from the solids in the disc, collide with each other, and migrate in the disc, while some of them get engulfed by planet. More than 2'000 individual simulations were run in a population synthesis framework, each one with a different dust-to-gas ratio, refilling timescale and number of initial seeds, chosen within appropriate limits (described in Sect. \ref{Methods}).

Our results show that the average satellite mass is below Galilean-masses, only 15\% of moon-systems being more massive than the Galilean system. The moonlets grow slowly, and most frequently, it takes $10^5$ years to reach Europa's mass. The migration rate is relatively slow, which means that only a few formed moons migrate all the way into the planet. In 10\% of the cases moons are engulfed by the planet, but only 1\% of the satellite-systems lose 1 Earth-mass or more into the planet. These lost moons and moonlets will contribute to the heavy element content of the giant planet's envelope. In all cases, the mass of the moons is mainly gained through accretion of solids, as oppose to collisions between moonlets. This highlights that the considered solid disc (in terms of dust-to-gas ratio, refilling timescale, etc.) might have a major role on the outcome of similar planet-formation models.

Checking for resonances, when all full-grown satellites and satellite-seeds are included in the analysis, 6.3\% of them are found to be in resonant-pairs. When only massive satellites are considered, 8.5\% of them are in resonance, with the most common configuration being (1:2) and (2:3).
If we consider only the Galilean-mass moons, 18\% of our systems have 3, 4, or 5 satellites. Smaller satellites are of course more numerous. 

We used a machine learning approach to compare our population with the Galilean system, applying the t-SNE algorithm, similarly to \cite{Alibert19}. This way, a "distance" between systems, meaning their similarity in the \textit{mass} $\times$ \textit{semi-major axis} space, has been defined and used to visualize and classify the outcomes. This approach demonstrates that the Galilean system is one possible outcome of the model and that the outcomes of the model depend almost solely on the parameter ($\Gamma = t_{\rm{ref}}/Z_{0.01}^{2.2}$). In particular, a Galilean-like system is produced when $\Gamma = 2\times10^4$ yr.

Our population of moons can be used to infer a potential exomoon population as well, around Jupiter-analogs. Therefore, we checked what fraction of our satellites could be detected with current and near-future instruments. Assuming the planet is already a transiting one, a sensitivity of $10^{-5}$ would allow to detect 14\% of moon-transits. To detect half of our population, a sensitivity of $2\times 10^{-6}$ would be necessary. Since the population is relatively small in mass, their transit-timing-variation reaches above 10 seconds only in 6.2\% of the case, posing a challenge to observe these moons.

We checked the difference made by having a 3D N-body model with 1D disc treatment, versus a fully 1D model (disc + dynamics). We found that the migration rates in the 3D case are lower, and hence the moons can grow to larger masses, and less frequently migrate all the way into the planet. Furthermore, in the fully 1D model, the scattering of satellites cannot be accounted for due to the limited dimensionality. In the 3D runs, there is a scattered, inner small moon population, that is missing from the fully 1D runs. 

We also examined the difference between 1D and 2D treatment of the circumplanetary disc in the planet formation model. The accretion proceeds differently, since only the solids within the feeding zone can be accreted (and therefore depleted) in the 2D case, while in the 1D case the entire azimuthal ring is available for accretion. This leads to the fact that in the 2D case initially the moons are smaller and accreting with a lower rate in comparison to the 1D. As the moonlets grow, their feeding zone enlarges too, even in the 2D case. The differential velocity within the feeding zone will then provide more solids available for the moon to accrete in the 2D case, and by the end of the formation, the results on the satellite masses between the 1D and 2D simulations are small, never more than 15\%.

\section*{Acknowledgements}

We would like to thank Richard Nelson, Gavin Coleman and Tom Hands for the very useful conversations we had, and also Thomas Ronnet for his precious feedback. We also thank Prasenjit Saha for his useful explanations of his N-boby integrator algorithms, and Darren Reed for his help with the computational resources. 

This work has been carried out within the framework of the National Centre of Competence in Research PlanetS supported by the Swiss National Science Foundation. The authors acknowledge the financial support of the SNSF. J.Sz. acknowledges the support from the Swiss National Science Foundation (SNSF) Ambizione grant PZ00P2\_174115.


\section*{Data availability}

The data underlying this article will be shared on reasonable request to the corresponding author.




\bibliographystyle{mnras}
\bibliography{bibliography} 




\appendix

\section{N-body integrator}\label{N-body integrator}

In the N-body integrator, satellites are considered as particles orbiting a central fixed particle with $M = 1M_J$. As the central particle is fixed, we consider it as an external potential, and all the sums in this Section will refer to the N satellites only.
The Hamiltonian of the system is then
\begin{equation}
    H = \sum_i^N\frac{p_i^2}{2m_i} - GM\sum_i^N\frac{m_i}{r_i} - G\sum_i^N\sum_{j>i}^N\frac{m_im_j}{r_{ij}}
\end{equation}
where $p_i$ is the momentum of the satellite $i$, $m_i$ is its mass, $r_i$ its distance to the centre and $r_{ij}$ is the distance between the particles $i$ and $j$.
This Hamiltonian is integrated following the algorithm shown by \cite{Saha94}. First, a leap-frog time-integrator is implemented, dividing the Hamiltonian into the Keplerian part (the drift) and the Interaction part (the kick), i.e.
\begin{equation}
\begin{split}
    &K = \sum_i^N\frac{p_i^2}{2m_i} - GM\sum_i^N\frac{m_i}{r_i}\\
    &I = - G\sum_i^N\sum_{j>i}^N\frac{m_im_j}{r_{ij}}
\end{split}
\end{equation}

This is more accurate than dividing the Hamiltonian into kinetic and potential energy, because the Interaction part is almost always a small correction to the Keplerian motion and this gives a more accurate method with the same cost. Analytic solutions for the Keplerian motion are well known and this model uses the same $f$ and
$g$ function method used in GENGA \citep{Grimm14} and described in \cite{Danby88}.

\subsection{Individual time-steps}\label{Individual time-steps}

In the model, the required precision is that any satellite needs at least 25 time-steps to complete one orbit, to keep consistency with the tests of \cite{Chambers99}. If the time-step were the same for all satellites, that would have been computationally too expensive due to the different orbital velocities of the inner and outer satellites within the disc. Instead, we split the Hamiltonian into multiple parts, each one related to one satellite
\begin{equation}
    \begin{split}
        K = \sum_i^N K_i\;\;\;\;\;&\text{with}\;\;\;\;\;K_i = \frac{p_i^2}{2m_i} - GM\frac{m_i}{r_i}\\
        I = \sum_i^N I_i\;\;\;\;\;&\text{with}\;\;\;\;\;I_i = - Gm_i\sum_{j>i}^N\frac{m_j}{r_{ij}}
    \end{split}
\end{equation}
This allows to solve the different parts of the Hamiltonian individually with different time-steps following the algorithm in Equation (14) in \cite{Saha94}.

The time-steps have to be chosen so that they are all proportional to a power of 2, i.e. $\Delta t_i =  2^{N_i}\Delta t_0$, where $\Delta t_0$ is chosen to be the orbital time of the inner orbit ($r=1R_J$) divided by 25, i.e. the required precision. $N_i$ is chosen in order to have $\Delta t_i$ as the minimum time-step with the required precision, i.e. less than the orbital time of the $i^{th}$ satellite divided by 25. In practice, if $T_i$ is the period of the $i^{th}$ particle, then $N_i = \rm{int}\left[\log_2\left(\frac{T_i}{25\Delta t_0} \right) \right]$. During the evolution of the system, satellites may vary their time-step because they move through the disc. In order to keep the code stable, a single satellite can only decrease its own time-step when migrating.
Finally, the algorithm provided by \cite{Saha94} needs the satellites to be ordered from the smallest to the larger $\Delta t_i$. This requires satellites to change also their index order in the code when migrating and changing their individual time-step.

\subsection{Close-encounters}\label{Close-encounters}

The algorithm works well as long as $I$ is a small correction of $K$. If the two become comparable, then the error of the algorithm, as given by the Baker–Campbell–Hausdorff formula \citep{Saha92, Saha94}, becomes one order of magnitude larger, and that happens when close encounters between satellites occur. This is managed implementing a close-encounter treatment from \cite{Chambers99}, following the example of GENGA \citep{Grimm14}. In case of close-encounters the Hamiltonian portions change to
\begin{equation}
    \begin{split}
        &K_i = \frac{p_i^2}{2m_i} - GM\frac{m_i}{r_i}- Gm_i\sum_{j>i}^N\frac{m_j}{r_{ij}}[1 - k_{ij}]\\
        &I_i = - Gm_i\sum_{j>i}^N\frac{m_j}{r_{ij}}k_{ij}
    \end{split}
\end{equation}
where $k_{ij}$ is a changeover function calculated for each pair of satellites. It has to be $1$ when satellites are far away and it needs to go to $0$ when they are coming closer. In particular
\begin{equation}
    \begin{split}
        k_{ij} = 
        \begin{cases}
            0\;\;\;&y<0\\
            y^3/(3y^2-3y+1)\;\;\;&0\le y<1\\
            1\;\;\;&y\ge1
        \end{cases}
    \end{split}
\end{equation}
with $y=\frac{r_{ij} - 0.1 r_{\text{crit}}}{0.9 r_{\text{crit}}}$,
where $r_{\text{crit}} = 3R_{\text{Hill},ij}$ and $R_{\text{Hill},ij}=\frac{a_i + a_j}{2} \left(\frac{m_i + m_j}{3M_J}\right)^{1/3}$ as suggested by \cite{Duncan98}. If close-encounters were involving also small satellites, a correction depending on velocities needs to be added for calculating $r_{\text{crit}}$, as explained in \cite{Chambers99}, but that is not the case in this model, as described below. The shape of $k_{ij}$ is chosen to be continuous and derivable twice in the whole domain.
Furthermore, we automatically assign $k_{ij}=1$ when at least one of the two satellites $i$ or $j$ has $M<M_{\text{threshold}}$, where the threshold mass is chosen to be $5\times10^{-6}M_J$. This is because close-encounters and collisions between small embryos are not significantly influencing the outcomes of this model.

When two or more satellites have $k_{ij} < 1$, the algorithm needs to change because $\{K_i, K_j\}\ne 0$ and the Keplerian parts of the Hamiltonian cannot be solved individually and analytically.
The Hamiltonian
\begin{equation}
    K_{\text{close-enc}} = \sum_i^N \frac{p_i^2}{2m_i} - GM\sum_i^N\frac{m_i}{r_i}- G\sum_i^N\sum_{j>i}^N\frac{m_im_j}{r_{ij}}[1 - k_{ij}]
\end{equation}
where $i, j$ are the indexes of satellites involved in the close-encounter, (i.e. the ones for which $k_{ij}<1$) can be solved, for example, implementing another leap-frog algorithm with a smaller time-step. As in \cite{Grimm14}, we decided to implement the Hermite Integrator with Ahmad-Cohen Scheme presented in \cite{Makino92}. This method is fourth-order because it computes the force and its derivative analytically, and then construct a third order polynomial interpolation. The method needs two parameters to compute the appropriate time-step at the beginning and during the integration. We chose these parameters to be $\eta = 0.02$ and $\eta_S = 0.01$, since these values have already been tested as the most efficient by \cite{Makino92}.

This close encounter treatment, as presented in \cite{Chambers99} and as used in \cite{Grimm14}, needs to be implemented in a democratic coordinate system, which consists of heliocentric (in our case planetocentric) positions and barycentric velocities. Since we have a central fixed particle (with infinite inertial mass) the democratic coordinates coincide with the Cartesian coordinate. 
Combining together individual time-steps and close-encounters is fundamental for this model in terms of performance and accuracy, but a synchronization problem may occur if two satellites that come closer do not have the same time-step, because it is not possible to integrate $K_{\text{close-enc}}$. If the two particles are synchronized, i.e. their internal clock is the same (see \citealt{Saha94} for details), there is no such issue. Then, the longer time-step is changed to the shorter time-step and the calculation can go on as described above. If that is no longer the case, then the code comes back to the last moment where the two satellites were synchronized, changes both the time-steps and starts over from there. In order to do that, the code is keeping in memory some back-up states of the system long enough to deal with these situations.

\section{N-body code validation}\label{N-body code validation}

In order to validate our N-body integrator, we tested and compared it with another code, i.e. GENGA \citep{Grimm14}, that uses similar methods for integrating the system's Hamiltonian and managing close-encounters (but no individual time-steps are implemented). Here we present the results of these tests.

First of all, we checked the energy conservation in our results. In order to do that, we produced 700 tests in which a variable number of satellites (between 10 and 20) were initialized with masses between $10^{-7}M_J$ and $10^{-4}M_J$. They were then let free to evolve, to have close encounters, to collide or even to be ejected. In these tests, no new satellites were allowed to form, since we compared them to some GENGA runs, that did not have this feature. For the same reason, the interaction with the gas was also removed. Our integration is considered to be good if, between each pair of consecutive collisions/ejections, energy is stable within a few percent, so that the energy evolution plot would look like a piece-wise step function. 

\begin{figure}
\includegraphics[width=\columnwidth]{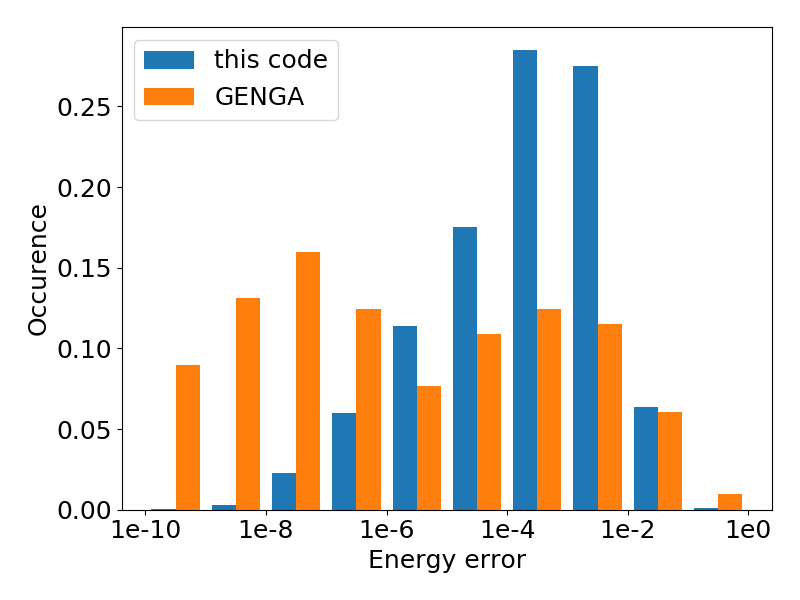}
\caption{The figure shows the energy errors measured in Genga and in our code. GENGA reaches much better accuracy for small satellites, but our code behaves well enough in massive cases.}\label{GlobalEnergyConservation}
\end{figure}

In order to check this, we fitted energy between any two collisions (happening in a time interval $\Delta t$) with linear functions in time $E = At + B$. For a perfect conservation, then $B=0$, hence we estimate the error as $\Delta E = \left|\frac{A\Delta t}{B}\right|$. All the $\Delta E$'s for any time interval are shown in Figure \ref{GlobalEnergyConservation}, where they are all compared with the same energy conservation values of GENGA. 

First of all, we notice that our code is able to conserve energy very well. Most of the intervals were conserving energy below $1\%$ and only a few cases, that involved more massive satellites in very tight close-encounters, have energy errors up to $10\%$. Compared to the GENGA's results, our code is handling these case a bit better, with GENGA having more cases with higher energy errors. On the other hand, "simpler" cases with smaller satellites in no particular configuration are handled much better by GENGA. This is due to the fact that our close-encounter treatment is enabled only for satellites with $M\ge 5\times 10^{-6} M_J$. This means that we do not reach extremely high energy conservation for smaller satellites, but this is not relevant for our results because small satellites are not the main constituents of the masses in our systems (although they are prevalent in number) and because migration would already cause satellites to change their energy.

\begin{figure}
\includegraphics[width=\columnwidth]{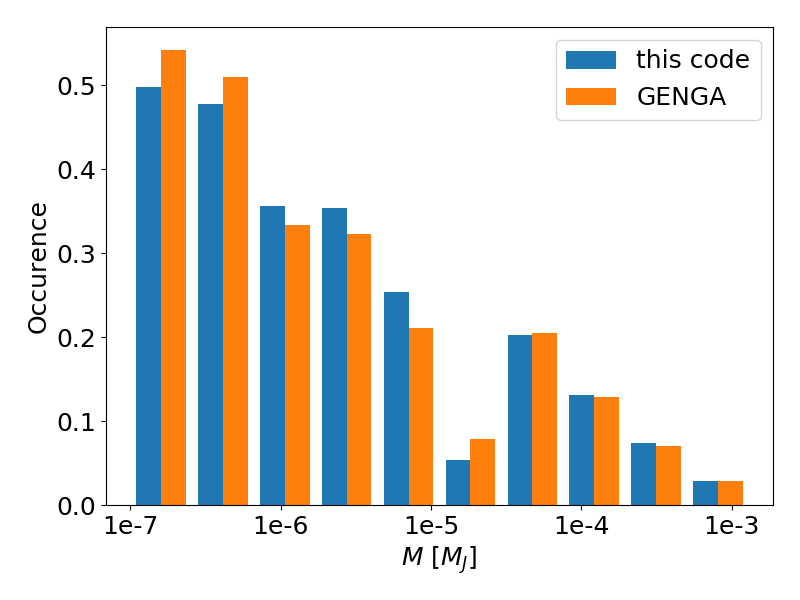}
\includegraphics[width=\columnwidth]{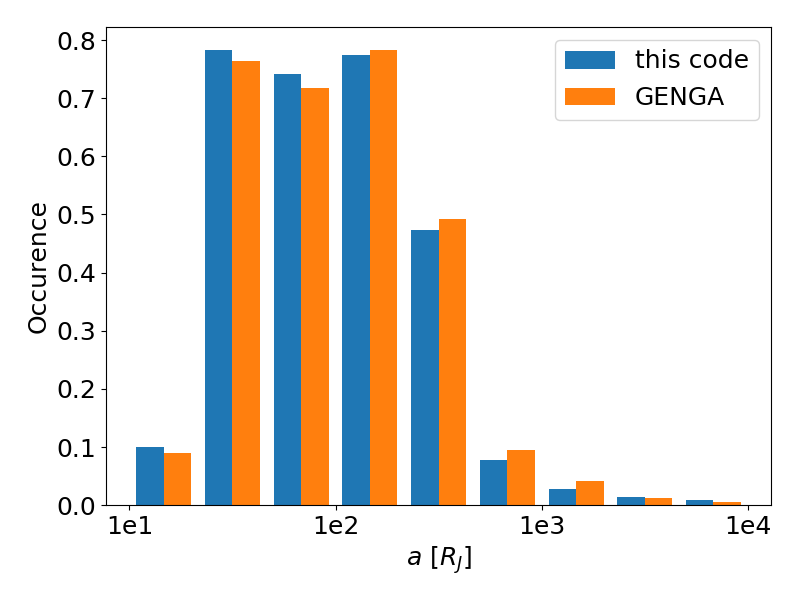}
\caption{The figures show the comparison of the mass and the semi-major axis distributions between our code and GENGA. The similarity of the results is a good evaluation of the behaviour of the code.}\label{Comparison}
\end{figure}

Another useful comparison between the two codes would be checking the average configuration of the results. Even if we started the tests with the same initial conditions, it is not reasonable to compare the outcomes one-to-one, because N-body systems are highly chaotic a small differences in the parameters and set-ups can lead to huge differences in the results. On the other hand, in Figure \ref{Comparison} we show a comparison between the masses and the semi-major axes of the final satellites.

In this case, the mass distribution is a good check on the collision frequency, as collisions are the only source of accretion in these tests. The two distributions are very similar and the small differences can be explained again by the close-encounter treatment. In our code, small satellites do not experience close-encounters, but they are considered as colliding more easily. This is why GENGA produces more small satellites, whereas the results almost coincide at higher masses. The semi-major axis distribution is also very similar, with GENGA pushing more satellites to the outer part of the system, probably because a better handling of close-encounters with small satellites.

\section{Randomized Dependence Coefficient}\label{Randomized Dependence Coefficient}

In Section \ref{Architecture of systems} an important parameter has been identified, i.e. $\Gamma = t_{\rm{ref}}/Z^{2.2}$, that is supposed to control the architecture of systems. Simulations with the same ratio between these two quantities will produce similar outcomes. The relation between them has been checked in Figure \ref{distances_galileans1}, \ref{distances_galileans2}, and \ref{TSNE_param}, but it can also be formally found thanks to a procedure known as Randomized Dependence Coefficient \citep{RDC, RDCpython}. The RDC is a measure of nonlinear dependence between random variables based on the Hirschfeld-Gebelein-Renyi Maximum Correlation Coefficient \citep{Hirschfeld35}. This coefficient will be 1 if two variables are correlated, even if non-linearly, 0 if they are not.

Given that the initial number of seeds does not influence the results significantly, the idea is that therefore the outcomes of the model would depend on some combination of the two other parameters. Mathematically, some quantity $Q$ that identifies the outcomes would be a function of this combination, i.e. $Q = f\left(t_{\rm{ref}}Z^{\beta}\right)$,
where Q could be for example the total mass of satellites.
Even better, if we refer to the framework presented in Section \ref{Architecture of systems}, this quantity could be the position $Y$ of systems on one axis as computed by the t-SNE algorithm. In fact, even if t-SNE has been used to represent systems in a 2D space, it is also able to reduce the dimensions further and represent systems on a 1D line. That means, in practice, unfolding and projecting the tube-shape in Figure \ref{TSNE_param} on one dimension.

\begin{figure}
\includegraphics[width=\columnwidth]{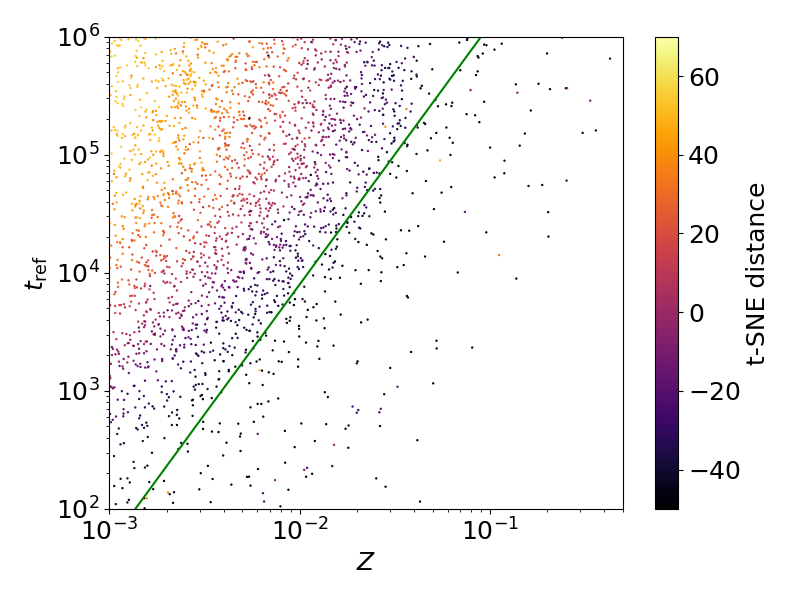}
\caption{The figure shows the position $Y$ of systems in a 1D projection got by a t-SNE approach as a function of the dust-to-gas ratio and the refilling timescale. Some correlation between this two parameters is easily visualized.}\label{TSNE_param_1D_0}
\end{figure}

\begin{figure}
\includegraphics[width=\columnwidth]{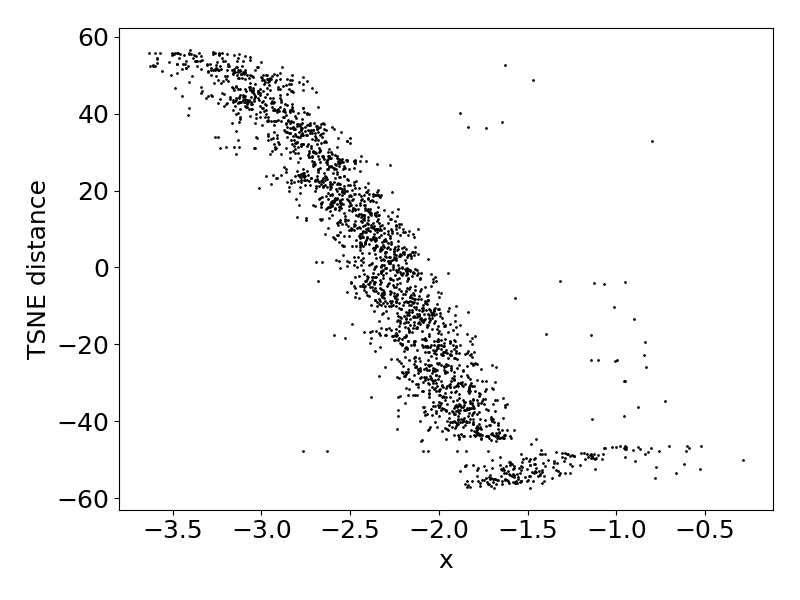}
\caption{The figure shows the position $Y$ of systems in a 1D projection got by a t-SNE approach as a function of the optimal $x = \log_{10}(Z/0.01) + \beta\log_{10}(t_{\rm{ref}}/\text{yr})$, i.e. with $\beta = -0.45$.}\label{TSNE_minimized_1D_0}
\end{figure}

If this intuition is true and we choose the right value for $\beta$, then the RDC between $X=t_{\rm{ref}}Z^{\beta}$ and $Y$ would be 1, or close to it. The procedure is then maximizing the RDC between $X$ and $Y$ while varying $\beta$, in order to find which combination of the two parameters is correlated the most with the results. It is already expected that a combination exists, given the results already shown in Section \ref{Architecture of systems}.
In order to do that, we studied the problem in a log-space, given that the results spanned over many orders of magnitude. Then the correlation can be examined between $x = \log_{10}X = \log_{10}Z + \beta\log_{10}t_{\rm{ref}}$ and the t-SNE 1D distance $Y$ ( Figure \ref{TSNE_param_1D_0} and \ref{TSNE_minimized_1D_0}). Figure \ref{TSNE_param_1D_0} shows the position $Y$ of systems as a function of the two parameters $Z$ and $t_{\rm{ref}}$, while Figure \ref{TSNE_minimized_1D_0} shows the distance $Y$ as a function of the $x$ for which the RDC has been maximized. In this case the algorithm has recognized a 95\% non-linear correlation with $\beta = -0.45$. This means that there is a non-linear correlation between the outcomes of the model and the parameter $\sim\frac{t_{\rm{ref}}}{Z^{2.2}} \propto \frac{t_{\rm{ref}}}{Z_{0.01}^{2.2}}$. 

\section{Observability model}\label{Observability model}

In Section \ref{Observability of system}, we used the concept of detection probability for a system around a Jupiter-like planet. First of all, anytime we considered this probability, we implied that that Jupiter-like planet is already detected and transiting. Otherwise the factor of $\frac{a_p}{R_{\star}} \simeq \frac{7\times10^7 m}{5.2AU}\sim10^{-4}$ has to be added in front of any probability \citep{Perryman11}.

Then, for each satellite $j$ in a given system $s$ we assigned a probability $p_j$. Once we have calculated those, the probability of detecting at least one satellite in that system would be
\begin{equation}\label{obs_probability}
    P_s=1-\sum_{j\in s}(1-p_j)
\end{equation}

There are two effects that contribute to $p_j$. The first one is trivial, i.e. we assign a probability of 1, if the transit depth of the satellite ($R_{\text{sat}}/R_{\star}$) is bigger than the sensitivity, while we assign 0 in the other case. Still, if a satellite is in principle detectable, we have to decrease the probability because of geometrical factors.
These factors actually represent the second effect, which has to do with the fact that, even if we detect the planet, a satellite could be not visible because it is hiding behind or in front of the planet or because it is not transiting in front of the star (i.e. having high inclination).

\begin{figure}
\includegraphics[width=\columnwidth]{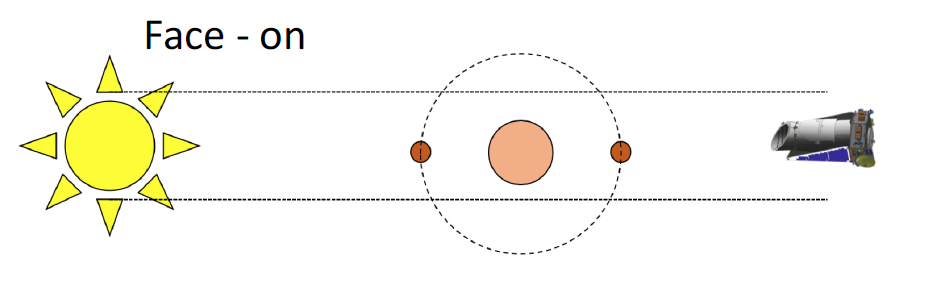}
\caption{Face-on configuration of the first geometrical effect. The satellite is transiting behind or in front of the planet during a portion of its orbit.}\label{transit1}
\end{figure}

In the first case, we refer to the scheme in Figure \ref{transit1}. There is a portion of the orbit of a satellite during which it is passing in front of or behind the planet, making it not visible with the transit method. As expected, the fraction of the orbit during which the satellite would be visible depends on the radius of the planet $R_p$, on the satellite semi-major axis ($a_s$) and on the inclination of the system $i$. Since the inclinations ($i$) and eccentricities ($e$) of satellites are usually low, in the calculations we assumed the planet and the satellite system to have the same inclination with the star-observer line and to have circular orbits.

In case of $i=\pi/2$, i.e. the case represented in Figure \ref{transit1}, the calculations are trivial and lead to $p_1 = 1 - \frac{2}{\pi}\arcsin\left(\frac{R_p}{a_s}\right)$, while taking into account a different inclination, we implemented the same formulae used for planetary transit duration \citep{Perryman11}:
\begin{equation}
    p_1(i) =
    \begin{cases}
    1 - \frac{2}{\pi} \arcsin \left(\sqrt{\frac{R_p^2/a_s^2 - \cos(i)^2}{1-\cos(i)^2}}\right)\;\;\;\;\; &|\cos(i)| < R_p/a_s\\
    1\;\;\;\;\; &|\cos(i)| \ge R_p/a_s
    \end{cases}
\end{equation}

\begin{figure}
\includegraphics[width=\columnwidth]{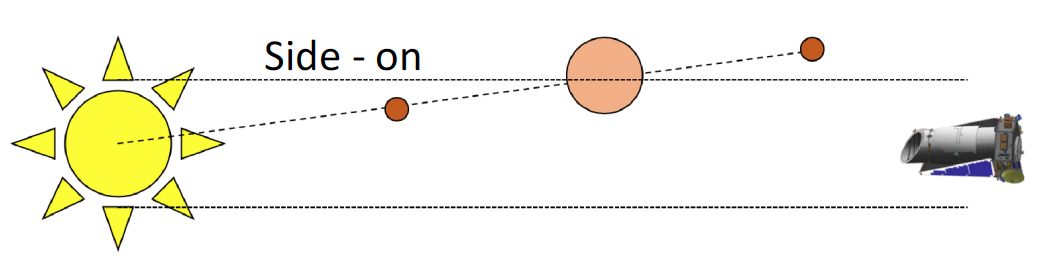}
\caption{Side-on scheme of the second geometrical effect. Even though the planet is transiting in front of the star, a satellite may spend part of its orbit out of the stellar disc.}\label{transit2}
\end{figure}

The second geometrical effect is shown in Figure \ref{transit2}. When the planet is visually close to the limb of the star, then a portion of the satellite orbit may be (visually) outside the stellar disc. With the planet transiting in front of the star, its distance from the edge is $R_{\star} - a_p \cos(i)$, that projected on the satellite orbit plane is $\frac{R_{\star} - a_p \cos(i)}{\cos(i)}$. With trigonometry calculations, we get that the fraction of the orbit during which the satellite is passing in front of the stellar disc, i.e. the second probability, is
\begin{equation}
    p_2(i) =
    \begin{cases}
    1 - \frac{1}{\pi} \arccos\left(\frac{R_{\star} - a_p \cos(i)}{a_s\cos(i)}\right)\;\;\;\;\; &|\cos(i)| > R_{\star}/(a_p +a_s)\\
    1\;\;\;\;\; &|\cos(i)| \le R_{\star}/(a_p +a_s)
    \end{cases}
\end{equation}

This two probabilities still depend on the inclination. If it is possible to detect a Jupiter-like planet then the inclination could be any number between $\arccos(R_s/a_p)$ and $\pi/2$, considering only positive inclinations. The final geometrical factor that we have to use has to be an integral over the inclination range, in particular
\begin{equation}
    \begin{split}
    p_{\text{geom}}&=\frac{\int_{\arccos(R_s/a_p)}^{\pi/2}p1(i)p2(i)\sin(i)di }{\int_{\arccos(R_s/a_p)}^{\pi/2}\sin(i)di }=\\ &=\frac{\int_0^{R_s/a_p} p1(\cos i)p2(\cos i) d\cos i}{R_s/a_p}
    \end{split}
\end{equation}

The integral is performed numerically in the analysis, the geometrical correction is applied to each satellite's detection probability and then the system observability is calculated via Equation \ref{obs_probability}.


\bsp	
\label{lastpage}
\end{document}